\documentclass[twocolumn,  twocolappendix]{aastex701}

\newcommand\aastex{AAS\TeX}

\shortauthors{Jheonn et al.}
\usepackage[utf8]{inputenc} % Required for inserting images
\usepackage{graphicx}
\usepackage{multirow}
\usepackage{appendix}
\usepackage{siunitx}
\usepackage{subfig}
\usepackage{comment}
\usepackage{tabularx}
\usepackage{amsmath}
\usepackage{textcomp}
\usepackage[flushleft]{threeparttable}
\newcolumntype{Y}{>{\centering\arraybackslash}X}

\begin{document}

\title{Template \aastex Article with Examples: 
v6.3.1\footnote{Released on March, 1st, 2021}}

\title{Refined classification of YSOs and AGB stars by IR magnitudes, colors, and time-domain analysis with machine learning}

\author[0009-0007-0657-3394]{Hyunwook Jheonn}
\email{jhw5023@snu.ac.kr}
\affiliation{Department of Physics and Astronomy, Seoul National University, 1 Gwanak-ro, Gwanak-gu, Seoul 08826, Republic of Korea}

\author[0000-0003-3119-2087]{Jeong-Eun Lee}
\email{lee.jeongeun@snu.ac.kr}
\affiliation{Department of Physics and Astronomy, Seoul National University, 1 Gwanak-ro, Gwanak-gu, Seoul 08826, Republic of Korea}
\affiliation{SNU Astronomy Research Center, Seoul National University, 1 Gwanak-ro, Gwanak-gu, Seoul 08826, Republic of Korea}

\author[0000-0003-4010-6611]{Jinho Lee}
\email{leejinho@snu.ac.kr}
\affiliation{Department of Electrical and Computer Engineering, Seoul National University, 1 Gwanak-ro, Gwanak-gu, Seoul 08826, Republic of Korea}

\author[0000-0001-6324-8482]{Seonjae Lee}
\email{sunjae627@snu.ac.kr}
\affiliation{Department of Physics and Astronomy, Seoul National University, 1 Gwanak-ro, Gwanak-gu, Seoul 08826, Republic of Korea}

\author[0000-0003-3130-7921]{Hyeyoon Lee}
\email{hylee817@snu.ac.kr}
\affiliation{Department of Electrical and Computer Engineering, Seoul National University, 1 Gwanak-ro, Gwanak-gu, Seoul 08826, Republic of Korea}

\author[0009-0002-1079-8178]{ShinGeon Kim}
\email{happyriosshs@snu.ac.kr}
\affiliation{Department of Electrical and Computer Engineering, Seoul National University, 1 Gwanak-ro, Gwanak-gu, Seoul 08826, Republic of Korea}

\author[0000-0003-1894-1880]{Carlos Contreras Pe\~na}
\email{}
\affiliation{Department of Physics and Astronomy, Seoul National University, 1 Gwanak-ro, Gwanak-gu, Seoul 08826, Republic of Korea}
\affiliation{Research Institute of Basic Sciences, Seoul National University, Seoul 08826, Republic of Korea}

\author[0000-0002-1408-7747]{Mi-Ryang Kim}
\email{mi.ryang@snu.ac.kr}
\affiliation{Department of Physics and Astronomy, Seoul National University, 1 Gwanak-ro, Gwanak-gu, Seoul 08826, Republic of Korea}

\correspondingauthor{Jeong-Eun Lee}
\email{lee.jeongeun@snu.ac.kr}

\begin{abstract}

We introduce a binary classification model, {\it the Double Filter Model}, utilizing various machine learning and deep learning methods to classify Young Stellar Objects (YSOs) and Asymptotic Giant Branch (AGB) stars. Since YSOs and AGB stars share similar infrared (IR) photometric characteristics due to comparable temperatures and the presence of circumstellar dust, distinguishing them is challenging and often leads to misclassification. While machine learning and deep learning techniques have helped reduce YSO-AGB misclassifications, achieving a reliable separation remains challenging. Given that YSOs and AGB stars exhibit distinct light curves resulting from different variability mechanisms, our Double Filter Model leverages light curve data to enhance classification accuracy. This approach uncovered YSOs and AGB stars that were misclassified in IR photometry and was validated against Taurus YSOs and spectroscopically confirmed AGB stars. We applied the model to the {\it Spitzer/IRAC Candidate YSO Catalog for the Inner Galactic Midplane} (SPICY) catalog for catalog refinement and identified potential AGB star contaminants.

\end{abstract}

\section{Introduction}

Young Stellar Objects (YSOs) are stars in the early stages of formation, embedded in their natal molecular clouds and still accreting material from their surrounding environments. They are generally classified into evolutionary stages—Class 0, I, II, and III—based on the shape of their spectral energy distributions (SEDs) and the presence of circumstellar material \citep{1987Lada, 1993Andre, 1994Greene}. Class 0 and I sources, called protostars, are deeply embedded protostars that are still undergoing significant accretion through a circumstellar disk, often accompanied by bipolar outflows and jets driven by magnetohydrodynamic processes \citep{2014Frank, 2016Bally}. As they evolve into Class II and III objects, their envelopes dissipate, revealing protoplanetary disks where planet formation occurs and disk evolution proceeds through accretion, grain growth, and disk clearing \citep{2011Williams, 2016Hartmann}. The SEDs of protostars and disk sources peak at infrared (IR) wavelengths because the emission from the central sources is reprocessed (i.e., absorbed and re-emitted) by the material in envelopes and disks.

% Introduction of AGB
Asymptotic Giant Branch (AGB) stars are post-main sequence stage stars, which low- to intermediate-mass stars (0.5 M$_\odot$ $<$ M $<$ 8 M$_\odot$) evolve into, before post-AGB star and white dwarf stages. Due to stellar pulsation, AGB stars lose most of their mass through stellar wind, forming circumstellar envelopes \citep{1996Habing, 2006Guandalini}. The existence of circumstellar envelopes causes AGB stars to exhibit IR excess. In addition, AGB stars exhibit an effective temperature of 2,000 to 4,000 K \citep{1988Vanderveen}, resulting in a peak luminosity in the IR wavelength.

% Classification in IR region
As a result, YSOs and AGB stars are readily detectable in the IR wavelengths. Given this, studies have focused on distinguishing YSOs and AGB stars through IR color-magnitude diagrams (CMD) and color-color diagrams (CCD) based on physical properties and statistical analysis \citep[e.g.,][]{2008Robit, 2009Gutermuth, 2012Koenig, 2014K&L}. These studies have demonstrated qualitative aspects in classification and remain widely used today. However, some sources, such as faint red AGB stars and bright blue YSOs, are difficult to distinguish from one another \citep{2014K&L}, although high-mass YSOs are more readily separated from AGB stars in CCD and CMD space owing to their substantially redder infrared colors \citep{2002Lumsden}. Because these confusing sources are difficult to distinguish by IR photometry alone, other methods or information are required to supplement the classification more precisely. Spectroscopy is one of the most reliable methods for identifying YSOs and AGB stars, yet the number of existing observations is limited.

% Machine Learning and Deep Learning 
With recent advances in machine learning (ML) and its subclass, deep learning, researchers have increasingly applied these methods to improve the classification of YSOs and AGB stars in IR photometry. For example, \citet{2019Marton} evaluated the performance of several ML techniques using Gaia {\it Data Release 2} (DR2) and the {\it Wide-field Infrared Survey Explorer} (WISE) data. \citet{2019Akras} employed a decision tree algorithm, while \citet{2021Kuhn} adopted Random Forest (RF), an advanced variant of the decision tree method. More recently, \citet{2023Laks} applied ensemble methods combining ML approaches to distinguish between YSOs and AGB stars, achieving classification accuracy above 90\% and reducing the level of cross-contamination between the YSO and AGB catalogs. However, these methods were unable to fully resolve contamination within the overlapping regions of CMDs and CCDs. \citet{2023Laks} observed that the classification probabilities dropped below 90 \% for sources located in these overlapped regions. This limitation highlights that even sophisticated ML algorithms struggle with precise classification, a task humans also struggle with. Therefore, additional features that exhibit distinct trends between YSOs and AGB stars are essential for further refinement in classification.

In this paper, we use light curve data from the {\it Near-Earth Objects Wide-field Infrared Survey Explorer} (NEOWISE) survey as an additional classification criterion to refine existing YSO and AGB catalogs, leveraging distinct trends in light curves between YSOs and AGB stars. YSOs accrete gas and dust from surrounding disks and envelopes irregularly \citep{2015Vorobyov, 2019Carlos, 2021LeeY}. This causes YSO variability to be stochastic or episodic \citep{2021Park, 2021LeeY}. High-mass YSOs also show similar variability trends \citep{2025Chen, 2025Kim}. Meanwhile, AGB stars show periodic light curves due to stellar pulsation \citep{2008Whitelock, 2021Lee}. 

Motivated by \citet{2021Lee}, the goal of this study is to identify hidden AGB contamination in the {\it Spitzer/IRAC Candidate YSO Catalog for the Inner Galactic Midplane} (SPICY) catalog \citep{2021Kuhn}, the largest YSO catalog currently available, using light curve information, and to provide a candidate list for future spectroscopic confirmation. While \citet{2021Lee} analyzed about 5,000 light curves individually to identify AGB candidates, it is not feasible to examine nearly 120,000 SPICY sources in the same way. Instead, we aim to develop a machine-learning method that leverages time-domain datasets to directly extract AGB contamination from the SPICY catalog, which was itself constructed using machine learning. In addition, we aim to develop models that utilize a broader range of machine-learning approaches on multi-wavelength photometric data, thereby complementing and extending the method employed to construct the SPICY catalog.

In Section \ref{sec2}, we explain the data and catalogs used in this work. Section \ref{sec3} introduces our ML classification model and its performance. Section \ref{sec4} presents the validation checks for our model using confirmed Taurus YSOs and spectroscopically studied AGB stars. In Section \ref{sec5}, we provide the model application to the SPICY catalog. Section \ref{sec6} discusses the targets reclassified by our model and the caveats of this work, and Section \ref{sec7}  summarizes this work with a conclusion.

\section{Data \& Catalogs} \label{sec2}

\begin{figure*}
    \centering
    \includegraphics[width=1\textwidth ]{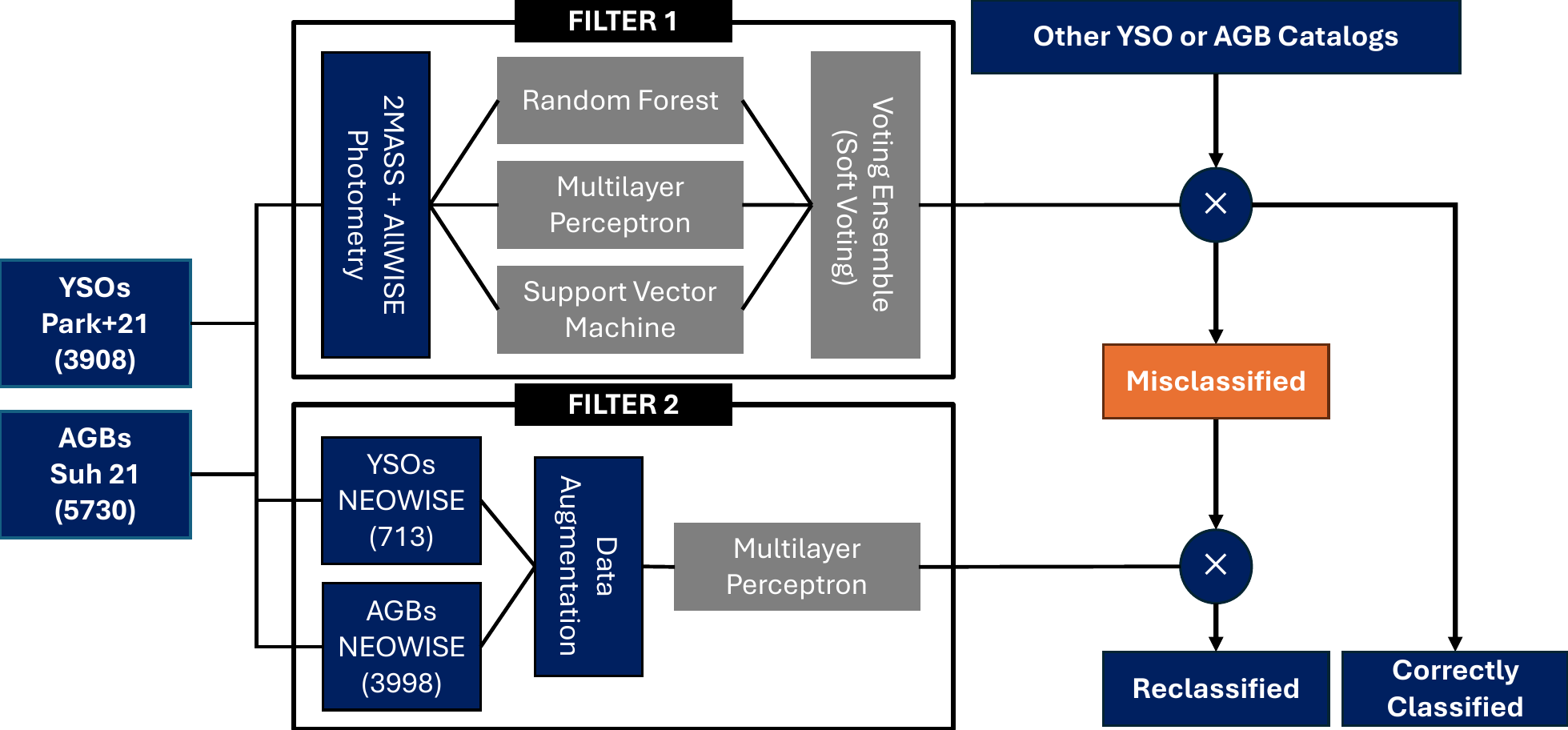}
    \caption{A visualized structure of the ``Double Filter classifier" model. Blue boxes represent the data or catalogs used for training and application, while gray boxes denote the ML methods. The voting ensemble in Filter 1 combines classifications into a single final result using a soft voting process. Data Augmentation in Filter 2 is an approach to increase the amount of data to mitigate overfitting caused by a lack of training data.}
    \label{figA}
\end{figure*}

As ML methods are data-driven, it is essential to select informative input
\emph{features} from which the model can effectively learn and generalize.
In supervised settings, ML typically requires at least thousands of labeled examples (and often more for complex models), with the exact number depending on the algorithm and model capacity \citep{2024Numtestset}.
Because our goal is to refine the classification of YSOs and AGB stars using near-IR (NIR) and mid-IR (MIR) photometry together with time-series information, we leverage
all-sky IR surveys (e.g., 2MASS, WISE/NEOWISE) that provide homogeneous measurements for thousands of sources.

\subsection{IR photometry and light curve}

%2MASS
The {\it Two Micron All-Sky Survey} (2MASS) is an all-sky survey carried out in the NIR J (1.25 $\mu$m), H (1.65 $\mu$m), and K (2.16 $\mu$m) bands between 1997 and 2001 \citep{20062MASS}. We selected 2MASS JHK data as one of our IR photometric inputs for ML analysis since 2MASS has provided extensive NIR photometry to date \citep[e.g.,][]{2008Robit, 2021Suh, 2023Laks}.

%WISE
We use the AllWISE catalog for the MIR photometric inputs. AllWISE is an all-sky survey performed by NASA's WISE space telescope \citep{2010WISE}. It consists of 4 MIR bands, W1 (3.4 $\mu$m), W2 (4.6 $\mu$m), W3 (12 $\mu$m) and W4 (22 $\mu$m). Most sources in the AllWISE catalog have 2MASS counterparts.

%NEOWISE
In September 2013, WISE was reactivated as the NEOWISE-reactivation mission \citep[NEOWISE-R,][]{2014Mainer}. NEOWISE performed all-sky observations using the W1 and W2 bands from 2013 to July 31, 2024. The telescope observed each region 10 to 20 times per epoch with a cadence of 6 months. We use NEOWISE 20-epoch data for YSOs and AGB stars, observed from 2013 to 2023, as inputs for the IR light curves, following the data processing method of \citet{2021Park}. 

\subsection{Catalogs used for Model Training}
\label{sec-train-cat}

%Catalog
We selected YSO and AGB star catalogs from \citet{2021Park} and \citet{2021Suh} for our training and test datasets of the ML model (hereafter referred to as P21 and S21). These are reliable catalogs, each containing over 5000 sources within our Galaxy. P21 compiled data from various nearby star-forming regions, Orion \citep{2012Megeath}, Taurus \citep{2019Esplin}, and Gould Belt \citep{2015Dunham}. However, the catalog in \citet{2015Dunham} contains a high level of AGB contamination \citep{2021Park}. \citet{2021Lee} identified AGB interlopers, which showed periodic light curves, within this dataset by detecting SiO maser lines. Therefore, we excluded these periodically variable sources that originated from the \citet{2015Dunham} catalog before model training. 

The S21 catalog has already undergone extensive filtering to remove contaminants and to confirm AGB stars through SED modeling, CMD and CCD analyses, light curves, and spatial distance verification \citep{2021Suh}. Thus, we adopted these sources in the catalog as true AGB stars.

We cross-matched the sources from each catalog against 2MASS, AllWISE, and NEOWISE. From the original P21 and S21 catalogs, 4023 YSOs and 5730 AGBs have complete data in all three surveys.
115 Taurus YSOs from P21 are preemptively set aside from these 4023 YSOs for future model validation in Section \ref{sec4}. As a result, 3908 YSOs and 5730 AGB stars comprise the training and test datasets for our model. For both datasets, we used 2MASS and ALLWISE data in magnitude units, and NEOWISE W1 and W2 data in Jansky (Jy).

\section{YSO-AGB classification model} \label{sec3}

\begin{figure*}
    \centering
    \includegraphics[width=1\textwidth ]{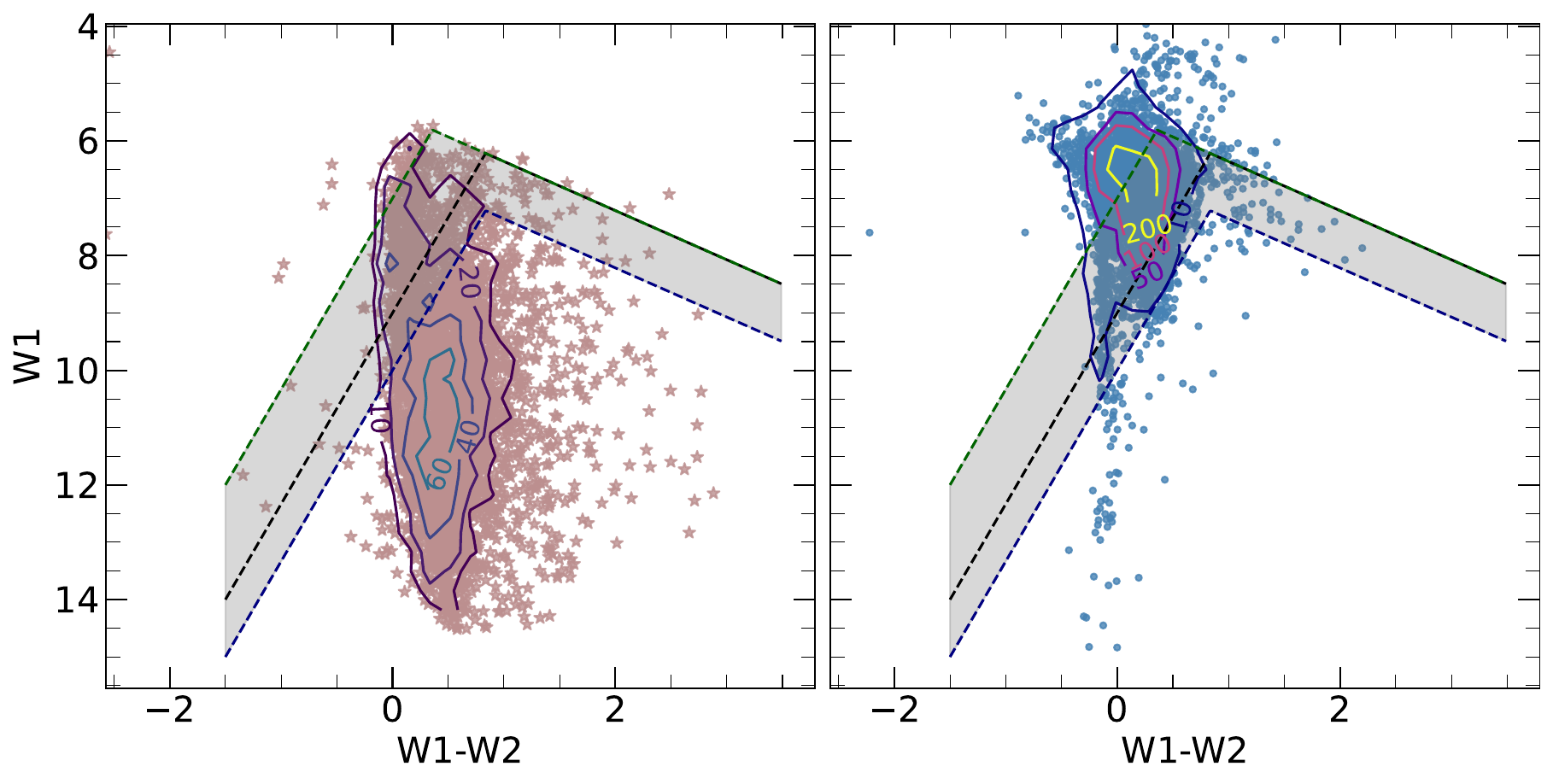}
    \caption{The P21 YSO ({\it left}) and S21 AGB star ({\it right}) distribution on W1-W2 vs. W1 CMD. The black-dashed lines in both graphs represent the division criterion for YSOs and AGB stars, as defined by \citet{2014K&L}. The gray-shaded regions are the overlapped regions defined by two empirically drawn lines (green and blue lines) among P21 and S21 catalogs based on the black dashed line.}
    \label{fig0A}
\end{figure*}

\subsection{Double Filter Classification Model}

{
We selected three widely used ML approaches for classification: Support Vector Machine (SVM) \citep{1995SVM}, Random Forest (RF) \citep{2001RF}, and Multilayer Perceptron (MLP) \citep{1991Murtagh}. %Feed-Forward Neural Network (FNN). 
SVM is well-suited for high-dimensional data and is often used in binary classification \citep{1995SVM}. RF, a popular method in YSO classification studies \citep{2019Marton, 2021Kuhn, 2023Laks}, is based on a bagging ensemble of many decision tree structures \citep{2001RF}.

MLP is a fundamental Neural Network method in ML. Given that our YSO and AGB star data consist of one-dimensional numerical values rather than image or video data, they do not require the use of complex and GPU-intensive ML architectures such as Convolutional Neural Network (CNN) \citep{2015CNN}, Recurrent Neural Network (RNN) \citep{2020RNN_funda}, or Transformer \citep{2017Transformer}. Instead, we applied Dropout \citep{2014Dropout} and Residual Connections \citep{2015He} within the MLP model to enhance performance and reduce overfitting \citep{2012Hinton, 2014Dropout}. For an overview of the ML methods, refer to \citet{2023OverallMLDL}.

We need to combine these ML results into an outcome, using an ensemble method \citep{2000Ensemble, 2021Ma_Ensemble_Review} and a two-stage classification method. The voting ensemble, one of the ensemble methods, is widely used in ML fields to improve classification accuracy \citep{2021Ma_Ensemble_Review}. This method concludes by either voting for the majority or averaging probabilities across multiple ML results. Two-stage classification uses two identical or different ML methods in sequence to refine or subclassify the classification (e.g., models of \citet{2022Dutta} and \citet{2023Laks}). We adopted these two methods for our model structure to improve classification performance and identify hidden contaminants within each other. 

Figure \ref{figA} shows the structure of our model. We named our model the {\it ``Double Filter model"}, which consists of two filters: ``Filter 1" and ``Filter 2". We use \texttt{scikit-learn (sklearn)} package \citep{2011scikit-learn} for SVM and RF, and \texttt{pytorch} package \citep{2019Pytorch} for MLP.
}

\subsubsection{Filter 1: Ensemble model with photometry data} \label{sec2-2-1}

Filter 1 is an integration of SVM, RF, and MLP trained with 2MASS and AllWISE photometry. The central concept for Filter 1 is to adopt conventional classification methodologies based on IR CCDs and CMDs, while using ML methods to follow a general YSO-AGB classification scheme. The inputs for Filter 1 consist of IR magnitudes (J, H, K, W1, W2, W3, and W4) and their corresponding colors, totaling 28 inputs. We organized 80 \% of the dataset from P21 (3908 YSOs) and S21 (5730 AGBs) as a training set, leaving 20 \% as a test set. 

The model uses the voting ensemble method to combine the results of each ML classifier into a final result. There are two combining cases: {\it hard voting} and {\it soft voting}. We chose soft voting because it provides both classification results and probabilities. A concrete explanation for the voting ensemble method is provided in Appendix \ref{appB}, along with the pseudo-code.

\subsubsection{Filter 2: MLP model with NEOWISE data} \label{sec2-2-2}

{
Another MLP serves as the main component of Filter 2. Twenty epochs of NEOWISE data, including W1 (Jy) and W2 (Jy) with their associated errors, form the basis of the input for the time-series analysis. The W1-W2 color for each epoch provides further potential insights for classification. We use $\Delta$W1/$\sigma$, $\Delta$W2/$\sigma$, and fractional amplitude (fAMP), defined in \citet{2021Park} as additional inputs of Filter 2.
$\Delta$W1/$\sigma$ and $\Delta$W2/$\sigma$ indicate the degree of amplitude compared to their mean magnitude uncertainty ($\sigma$) of W1 and W2, calculated as 
\begin{align}
\label{eq1-1}
\tag{1-1}
    \Delta W1 / \sigma &= \frac{W1_{mag, max} -W1_{mag, min}}{\frac{1}{20}\sum_{i=1}^{20}{\sigma_{W1_i}}}\\
\label{eq1-2}
\tag{1-2}
    \Delta W2 / \sigma &= \frac{W2_{mag, max} -W2_{mag, min}}{\frac{1}{20}\sum_{i=1}^{20}{\sigma_{W2_i}}}
\end{align}
The period of variability and fAMP calculated with the Lomb-Scargle periodogram \citep{1976Lomb, 1989Scargle} also provide significant clues in classification. fAMP is calculated as
\begin{align}
\label{eq2}
\tag{2}
    fAMP = \frac{Amplitude}{\overline{Flux}}
\end{align}

According to \citet{2021Park}, periodic YSOs and AGB stars exhibit different characteristics on fAMP and period: YSOs have fAMP lower than 0.15, while AGB stars show fAMP greater than 0.15, with a period range from 100 to 1000 days \citep{2008Whitelock, 2018Hofner}. These quantities are meaningful only for sources with periodic light curves. We therefore use fAMP and period only for sources with a false alarm probability (FAP) $<0.01$; for all others, we set these values to zero for modeling convenience.

% along with the NEOWISE data. 
We define an “overlap region” in the W1–W2 vs. W1 CMD where P21 and S21 sources intersect (gray shading in Figure~\ref{fig0A}). Guided by the criterion of \citet{2014K&L} (black dashed line in Figure~\ref{fig0A}), we draw two empirical boundaries (blue and green dashed lines) that delimit the zone in which YSOs and AGB stars overlap.

The training and test datasets for Filter 2 were restricted to this overlapped region to make the model focus on the light curve trends within this region. 
This magnitude–color restriction substantially reduces the available training data to 713 YSOs and 3,998 AGB stars, which increases the risk of overfitting during training \citep{2017DataAugmentation}.

To address this, we applied `data augmentation' to expand the training set. Data augmentation amplifies the number of datasets based on true data \citep{2017DataAugmentation} by inserting artificial Gaussian noise and scaling the data. This process reduced the instability in the Filter 2 training and stabilized its performance.

In total, we use 104 features per source as inputs of Filter 2: W1, W1 errors, W2, W2 errors, and W1-W2 for each of 20 epochs, as well as $\Delta$W1/$\sigma$, $\Delta$W2/$\sigma$, period, and fAMP. The 713 YSOs and 3998 AGB stars are divided into training and test datasets, with a ratio of 80 \% and 20 \%, before replenishing insufficient training data with data augmentation up to 6303 and 6364, respectively.  }

\subsection{Model Performances} \label{sec3-2}

\begin{table*}[]
\centering
\caption{The Accuracy, Precision, Recall, and F1-score of each ML method in Figure \ref{figB}.}
\label{tab3}
\begin{threeparttable}
\begin{tabular*}{\textwidth}{@{\extracolsep{\fill}} c *{5}{S[table-format=1.3]}}
\toprule
\multirow{2}{*}{\textbf{Evaluation Metrics}} & \multicolumn{3}{c}{\textbf{Filter 1}} & \textbf{Filter 2} \\ \cline{2-4} \cline{5-5}
                   & \textbf{SVM} &  \textbf{MLP(Phot)} &  \textbf{RF} & \textbf{MLP (NEOWISE)} \\ \hline
\textbf{Accuracy}  & 97.0 \%      &  96.6 \%            &  97.0 \%          & 94.5 \%              \\ \hline
\textbf{Precision} & 97.3 \%      &  96.7 \%            &  97.0 \%          & 77.4 \%              \\
\textbf{Recall}    & 95.2 \%      &  94.7 \%            &  95.5 \%          & 87.9 \%              \\ \hline
\textbf{F1-Score}  & 96.3 \%      &  95.7 \%            &  96.2 \%          & 82.3 \%              \\ \hline
\\
\end{tabular*}
\end{threeparttable} 
\end{table*}

\begin{table}
\centering
\caption{The Accuracy, Precision, Recall, and F1-score of Filter 1 and 2 in Figure \ref{figD}, and the ``Double Filter model" test, which is the combination of both filters.}
\label{tab4}
\begin{tabular}{cccc}
\toprule
\textbf{}          & \textbf{Filter 1} & \textbf{Filter 2} & \textbf{Double Filter Model}  \\ \hline
\textbf{Accuracy}  & 97.1   \%         & 76.4   \%      & 99.3 \%        \\ \hline
\textbf{Precision} & 97.5   \%         & 89.7   \%      & 99.6 \%         \\
\textbf{Recall}    & 95.3   \%         & 72.2   \%      & 98.7 \%        \\ \hline
\textbf{F1-Score}  & 96.4   \%         & 80.0   \%      & 99.1 \%        \\ \hline
\end{tabular}
\end{table}

\begin{figure*}
    \centering
    \includegraphics[width=0.24\textwidth ]{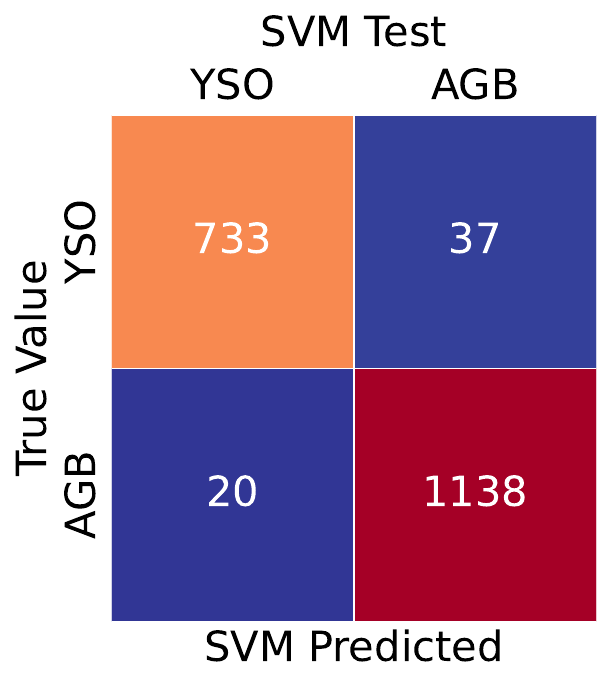}
    \includegraphics[width=0.24\textwidth ]{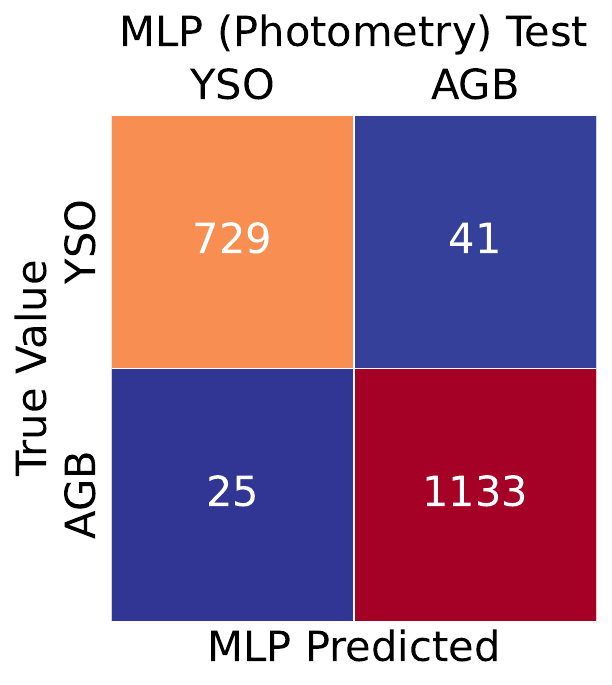}
    \includegraphics[width=0.24\textwidth ]{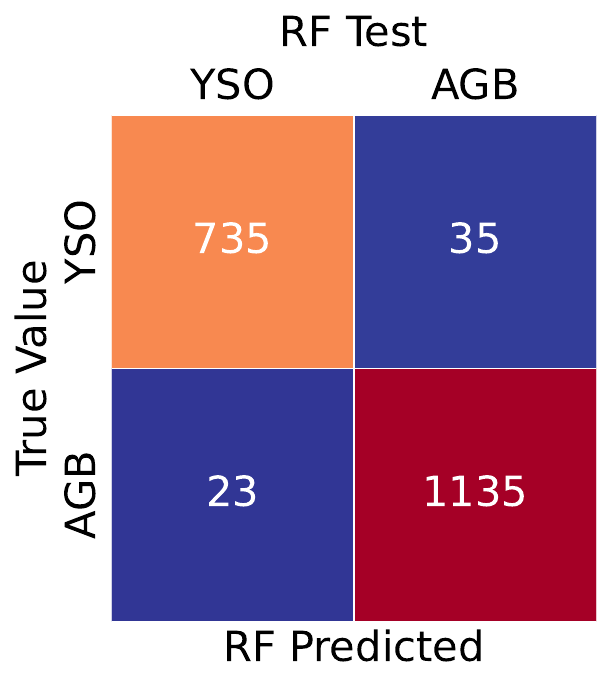}
    \includegraphics[width=0.24\textwidth ]{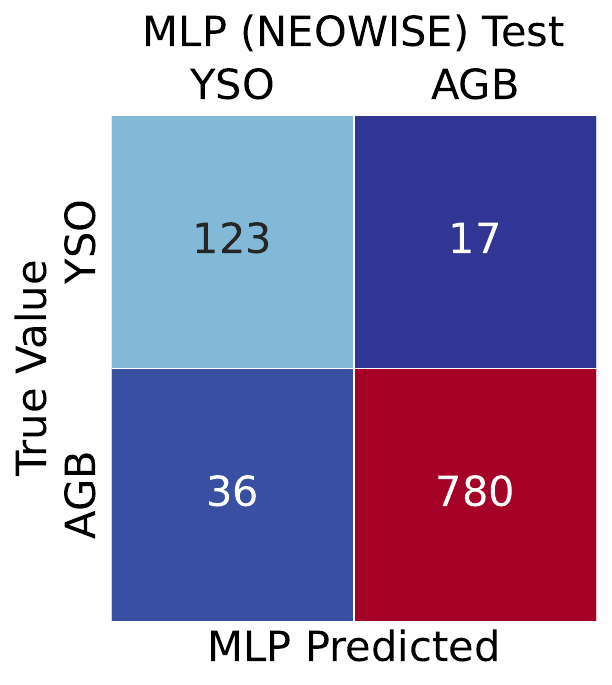}
    
    \caption{The confusion matrices of the test results for the three methods (SVM, RF, MLP (Photometry)) in Filter 1 and NEOWISE trained MLP, which is the base component of Filter 2.}
    \label{figB}
\end{figure*}

\begin{figure}
    \centering
    \includegraphics[width=0.45\textwidth ]{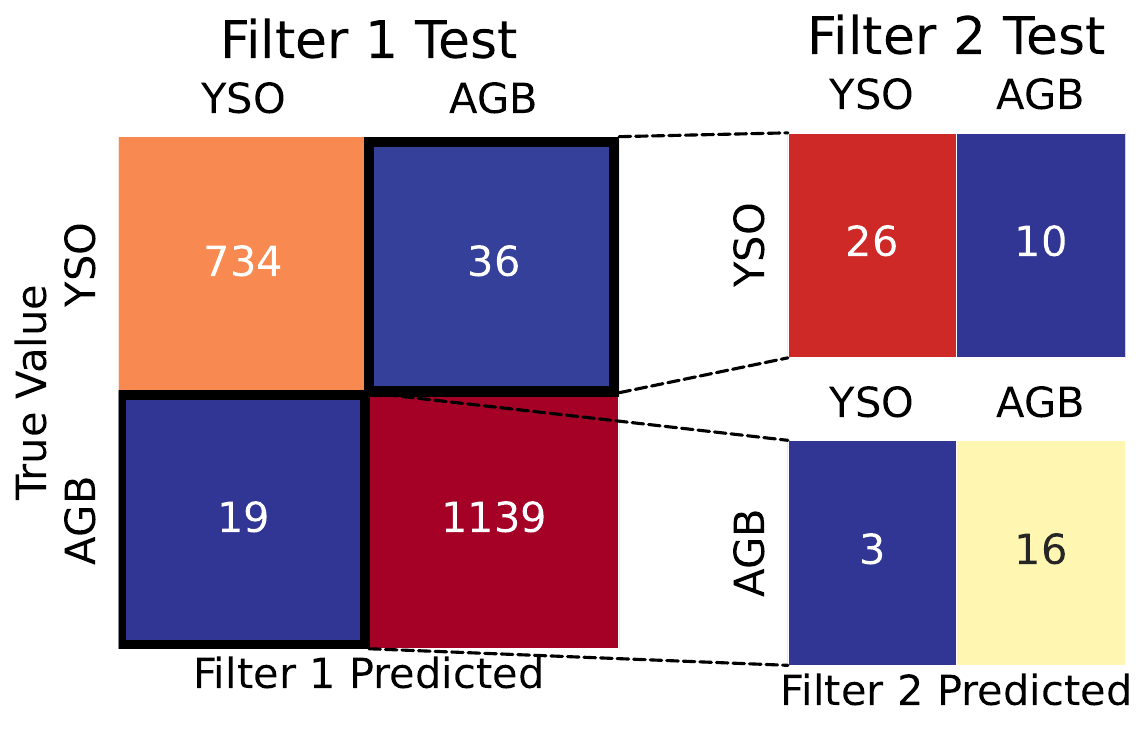}
    \caption{The confusion matrices of the Filter 
    1 and Filter 2 test results. Test data in Filter 2 are the misclassified YSOs (36) and AGBs (19) from Filter 1.}
    \label{figD}
\end{figure}

{
We used the 20 \% test data to assess its performance. Figure \ref{figB} shows the outcomes of the individual method in Filter 1 and Filter 2, represented as confusion matrices. The confusion matrix provides the classification results as True positive (TP), True negative (TN), False positive (FP), and False negative (FN). Each corresponds to YSOs classified as YSOs, AGB stars classified as AGB stars, AGB stars misclassified as YSOs, and YSOs misclassified as AGB stars in this paper. The accuracy, precision, recall, and F1-score values are the metrics for evaluating ML performance \citep{2020Grandini}. Each metric is represented by
\begin{align*}
    Accuracy &= \frac{TP+TN}{TP+TN+FP+FN}\\ 
    Precision &= \frac{TP}{TP+FP}\\
    Recall &= \frac{TP}{TP+FN}\\ 
    F1{\text -}score &= \frac{2\times Precision\times Recall}{Precision + Recall}   
\end{align*}

Filter 1 achieved more than 95 \% in accuracy and F1-scores (SVM, MLP(Phot), and RF in Table \ref{tab3}), while Filter 2 achieved approximately 88 \% (average of the accuracy and F1-score of MLP (NEOWISE) in Table \ref{tab3}). The integrated test result of Filter 1 and Filter 2 is represented in Figure \ref{figD}, following the classification process illustrated in Figure \ref{figA}. About 76 \% of misclassified YSOs and AGB stars by Filter 1 returned to their group after applying Filter 2 (Table \ref{tab4}). In total, about 99 \% of the test data were correctly classified by the Double Filter model. These results suggest that Filter 2 complements Filter 1, even though Filter 2 alone yields lower accuracy and F1-score than Filter 1, likely because of its limited wavelength information.

Although we assume the training and test datasets are uncontaminated, any sources misclassified in the final results warrant further investigation. We examine these sources in Section \ref{sec6} in detail to reveal their nature.}

\section{Model Validation} \label{sec4}

\begin{table}[]
\centering
\caption{The validation result of the model using 115 Taurus YSOs.}
\label{tab4-2}
    \begin{threeparttable}
        \begin{tabular*}{\columnwidth}{@{\extracolsep{\fill}} c *{4}{S[table-format=1.3]}}
            \toprule
            \textbf{}          & \textbf{confirmed YSOs}              & \multicolumn{2}{c}{\textbf{misclassified AGBs}} \\ \hline
            \textbf{Filter 1}  & {107}                               & \multicolumn{2}{c}{\textcolor{red}{\textbf{8}\tnote{a}}}     \\
            \textbf{Filter 2}  & \textcolor{blue}{\textbf{8}\tnote{b}}       & \multicolumn{2}{c}{\textcolor{blue}{\textbf{0}\tnote{b}}}   \\ \hline
            \textbf{Total}     & \textbf{115}                       & \multicolumn{2}{c}{\textbf{0}}                \\ \hline
        \end{tabular*} \vspace{5pt}
        \textbf{Notes.}
        \begin{tablenotes}[flushleft]\footnotesize
        \item[a] Misclassified as AGB stars by Filter 1.
        \item[b] Reclassified from [a] by Filter 2. All 8 were retrieved to YSOs.
        \end{tablenotes}
    \end{threeparttable}%
\end{table}%

\begin{table}[]
\centering
\begin{threeparttable}%
\caption{The validation result of the model using 85 validation AGB stars. }
\label{tab4-1}
\begin{tabular*}{\columnwidth}{@{\extracolsep{\fill}} c *{4}{S[table-format=1.3]}}
\toprule
\textbf{} & \textbf{misclassified YSOs} & \textbf{confirmed AGBs} \\ \hline
\textbf{Filter 1}  & \textcolor{red}{\textbf{45}\tnote{a}}           & {40}           \\
\textbf{Filter 2}  & \textcolor{blue}{\textbf{10}\tnote{b}}            & \textcolor{blue}{\textbf{35}\tnote{b}}          \\ \hline
\textbf{Total}     & \textbf{10}   & \textbf{75}           \\ \hline
\end{tabular*}\vspace{5pt}
    \textbf{Notes.}
        \begin{tablenotes}[flushleft] \footnotesize
            \item[a] Misclassified as YSOs by Filter 1.
            \item[b] Reclassified from [a] by Filter 2. 10 are the final misclassified YSOs due to the model error.
        \end{tablenotes}
        
\end{threeparttable}%
\end{table}

\begin{table*}[]
\centering
\caption{The variable types of validation AGB stars misclassified as YSOs by Filter 1, reclassified AGB stars, and remaining misclassified YSOs by Filter 2. The variability types are from \citet{2021Park} and \citet{2024LeeS}, based on the Lomb-Scargle periodogram.}
\label{tab2-1}
\begin{tabularx}{\textwidth}{YYYYY}
\toprule
\multicolumn{2}{c}{\bf Variability Type}                &  {\bf Filter   1 (YSO)}       & {\bf Filter   2 (AGB)} & {\bf Filter   2 (YSO)} \\ \hline
\multicolumn{1}{c}{\multirow{3}{*}{Secular}}    & Periodic  & 30  (66.67 \%)   & 29  (82.86 \%)   & 1    (10.00 \%)   \\
\multicolumn{1}{c}{}                            & Curved    & 1   (2.22 \%)    & 1   (2.86 \%)    & 0    (0.00 \%)  \\
\multicolumn{1}{c}{}                            & Linear    & 0   (0.00 \%)    & 0   (0.00 \%)    & 0    (0.00 \%)    \\ \hline
\multicolumn{1}{c}{\multirow{3}{*}{Stochastic}} & Burst     & 0   (0.00 \%)    & 0   (0.00 \%)    & 0    (0.00 \%)    \\
\multicolumn{1}{c}{}                            & Drop      & 0   (0.00 \%)    & 0   (0.00 \%)    & 0    (0.00 \%)    \\
\multicolumn{1}{c}{}                            & Irregular & 5   (11.11 \%)    & 4   (11.42 \%)   & 1    (10.00 \%)    \\ \hline
\multicolumn{2}{c}{Non-variable}                         & 9   (20.00 \%)   & 1   (2.86 \%)    & 8    (80.00 \%)  \\ \hline
\multicolumn{2}{c}{\bf Total}                            & {\bf 45}    &  {\bf 35}    & {\bf 10}    \\ \hline

\end{tabularx}
\end{table*}

\begin{figure*}
    \centering
    \includegraphics[width=0.45\textwidth ]{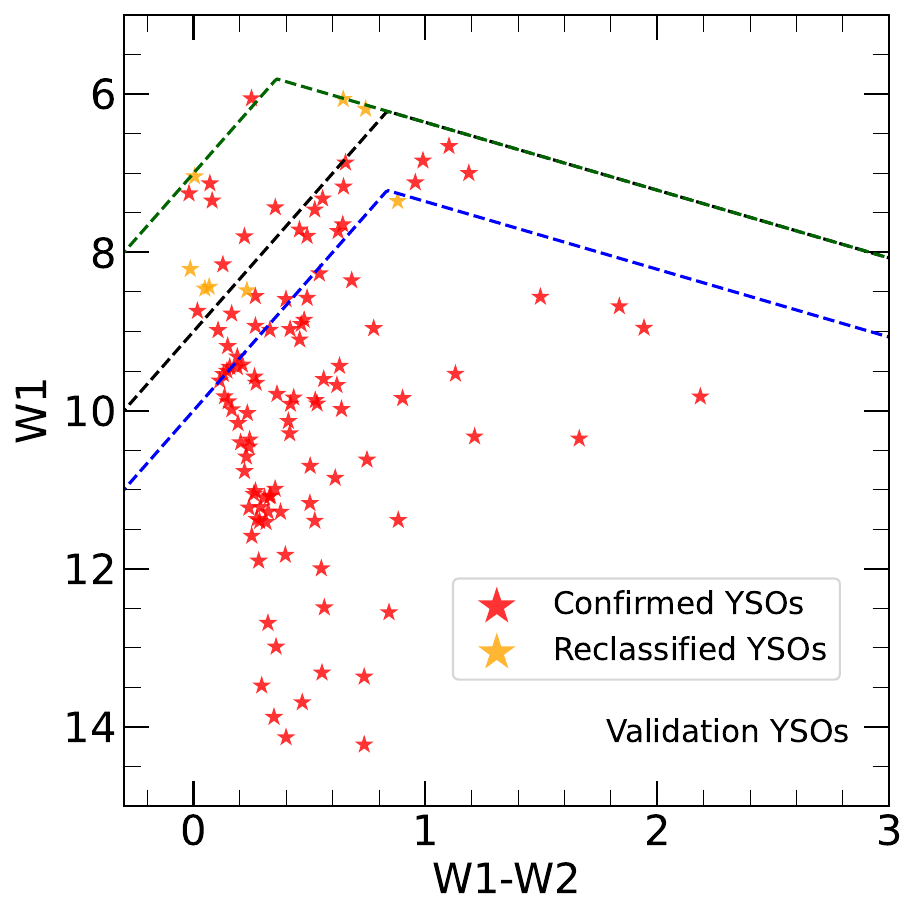}
    \includegraphics[width=0.45\textwidth ]{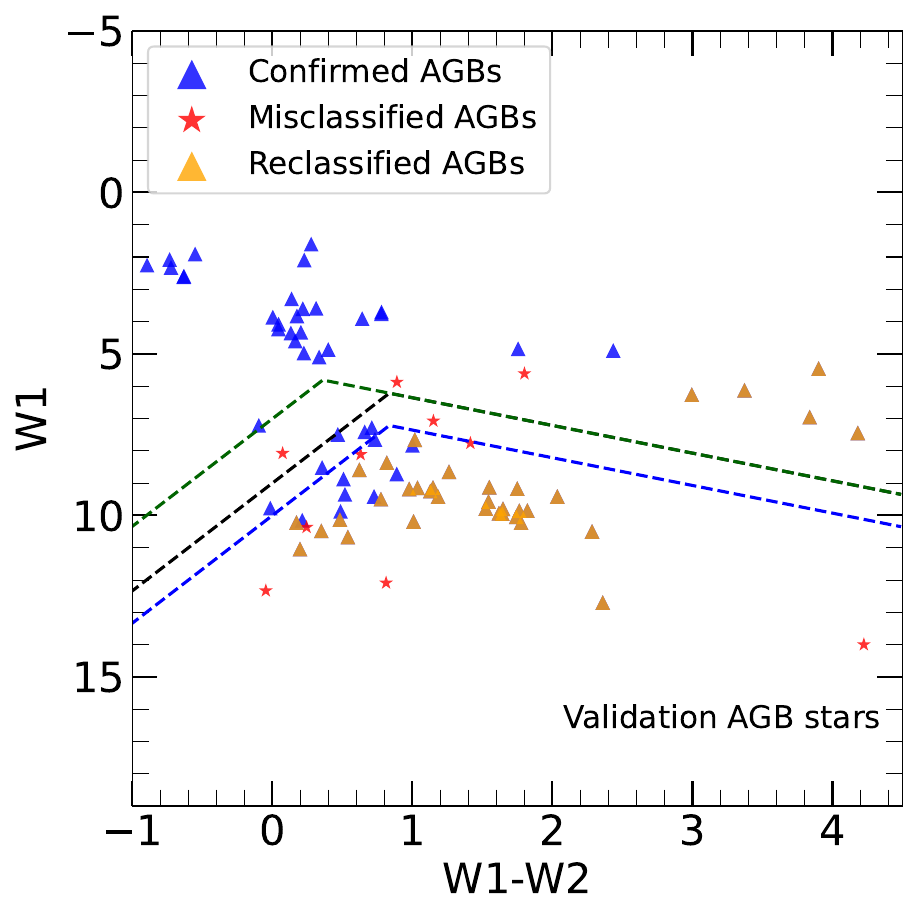}
    \caption{\textit{Left}: The W1-W2 vs. W1 CMD of 115 Taurus YSOs. Red stars are confirmed YSOs by Filter 1, and orange stars are eight reclassified YSOs by Filter 2. \textit{Right}: The same CMD for 85 validation AGB stars. Blue triangles are the AGB stars confirmed by Filter 1, yellow triangles are reclassified AGB stars by Filter 2, and red stars are the remaining misclassified AGB stars. The dashed lines are those in Figure \ref{fig0A}.}
    \label{figCC}
\end{figure*}

{The test results mentioned in Section \ref{sec3-2} proved that the performance of the Double Filter model is robust. However, the model still requires an additional validation process to support its ability. Here, we present two validation test results: one with 115 Taurus YSOs and the other with 85 spectroscopically confirmed AGB stars (hereafter referred to as validation YSOs and validation AGB stars). These targets are not included in either the training or the test datasets.}

\subsection{Validation YSOs: Taurus YSOs}

{
YSOs in the Taurus Molecular Cloud are the most studied ``known" YSOs, as the cloud is one of the nearest clouds to the Earth \citep{2018Galli, 2019Esplin}. Thus, it is reliable to use Taurus YSOs as a model validation. Because P21 contains all Taurus YSOs, we preemptively selected 115 Taurus YSOs before model training, as mentioned in Section \ref{sec-train-cat}. This process enables our model to continually learn the features of Taurus YSOs, test the model, and validate its performance without duplicating the data sets.  

The left panel of Figure \ref{figCC} shows the W1-W2 vs. W1 CMD of 115 Taurus YSOs, and Table \ref{tab4-2} shows the validation result of the model with these 115 Taurus YSOs. According to Filter 1, 107 of 115 are confirmed as YSOs with magnitudes and colors alone, while 8 YSOs are misclassified as AGB stars. These 8 misclassified YSOs are pure Filter 1 error, because they are ``known" YSOs \citep{2019Esplin}. Filter 2, however, returned all eight misclassified sources to the YSOs, thereby supporting the model's performance.
}

\subsection{Validation AGBs: Spectroscopically studied AGBs}

{ 
The validation AGB stars are the combined AGB catalog of three studies, \citet{1999VanLoon}, \citet{2008Raman}, and \citet{2024Tatarnikov}. We used 85 AGB stars containing 2MASS, AllWISE, and 20 epochs of NEOWISE data. Given that 47 AGB stars are from the Large Magellanic Cloud (LMC) \citep{1999VanLoon}, %they are darker and redder than AGB stars in our Galaxy due to different environments of gases, dusts, and metallicity in LMC \citep{2020Suh}. Since the model has never encountered AGB stars outside our Galaxy before, 
it is meaningful to determine whether the model classifies these LMC AGB stars as such. 

Table \ref{tab4-1} shows the results of the validation AGBs classified by Filter 1 and Filter 2. Filter 1 classified 45 of 85 sources ($\sim$53\%) as YSOs. LMC AGB stars are mostly responsible for this unexpected result, since 33 of 45 misclassified AGB stars were revealed as AGB stars from LMC. The right graph in Figure \ref{figCC} supports that these misclassified AGB stars by Filter 1 (yellow triangles and red stars) have higher magnitude and redder W1-W2 color. However, when these sources were tested with Filter 2, 35 were reclassified as AGB stars, leaving 10 still misclassified as YSOs. 

Table \ref{tab2-1} shows the variable types of 45 misclassified AGB stars by Filter 1, 35 reclassified AGB stars, and the remaining 10 sources misclassified as YSOs by Filter 2. Of the 45 misclassified as YSOs by Filter 1, 29 periodic AGB stars were retrieved as AGB by Filter 2. One periodic and 9 irregular or non-variable AGB stars were still regarded as YSOs, of which 6 were LMC AGB stars. In total, 75 out of 85 AGB stars were successfully classified as AGB stars, showing an accuracy of about $\sim$88.2 \%. Although Filter 2 has no experience with AGB stars from outer galaxies during training, it correctly classified 27 of 33 LMC AGB stars misclassified by Filter 1, which is confusing in the magnitude and color regime. Therefore, the results above support our argument that Filter 2 scrutinizes the light curve trend and returns misclassified sources to their original group.

As a result, the validation process with Taurus YSOs and spectroscopically confirmed AGB stars shows that our model achieves better performance than methods using only single-epoch IR photometry magnitude and color inputs.
}

\section{Refinement of the SPICY catalog} \label{sec5}
{
The Double Filter model demonstrated its effectiveness through model testing and validation. Therefore, we apply the model to the biggest YSO catalog to provide its original purpose: catalog refinement. Since the model is a binary classification model, only YSOs and/or AGB catalogs are applicable; the SPICY \citep{2021Kuhn} satisfies all these conditions.
}

\subsection{Application of Double Filter Model to the SPICY}

\begin{figure*}
    \centering
    \includegraphics[width=0.9\textwidth]{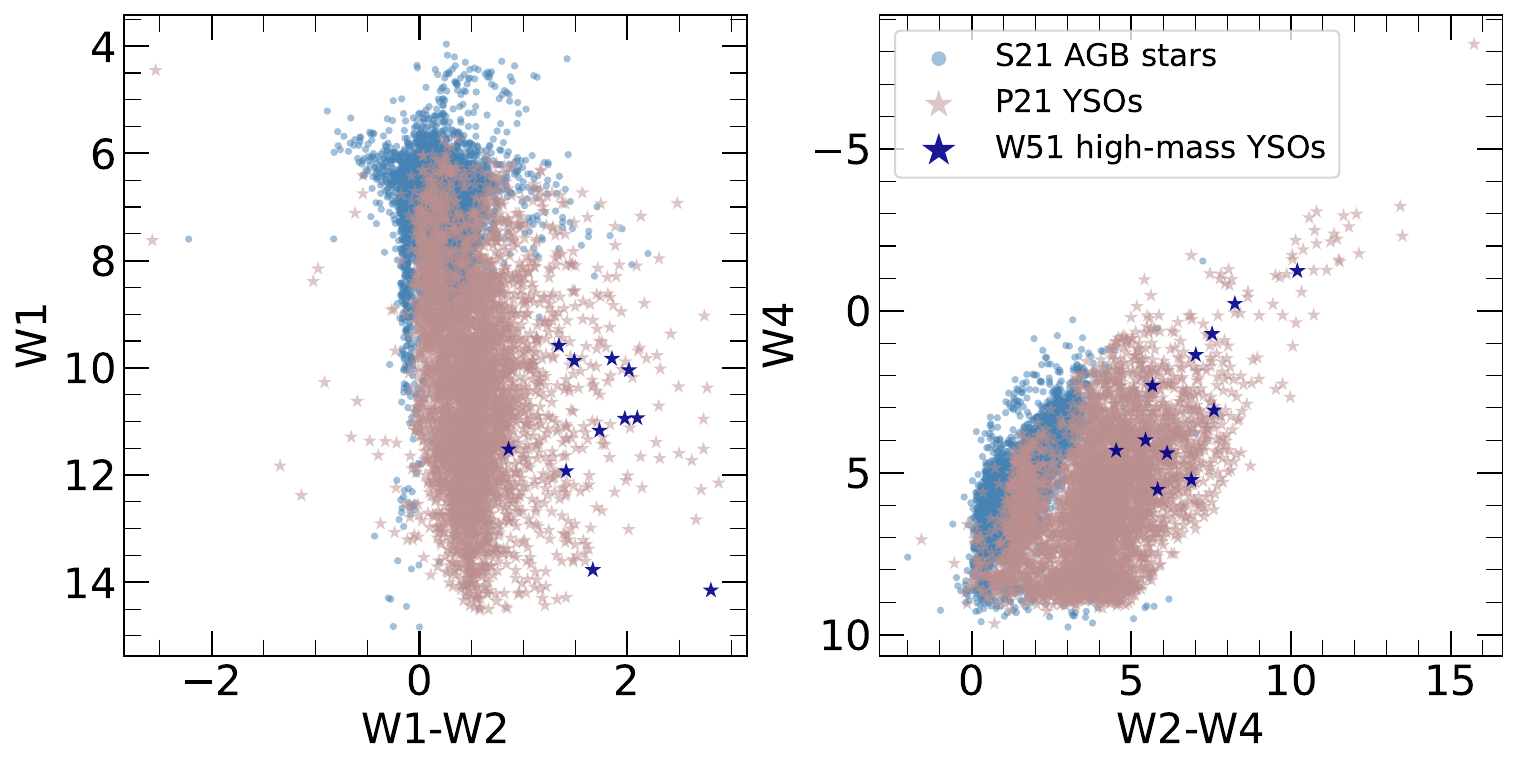}
    \caption{The W1-W2 vs W1 CMD ({\it left}) and W2-W4 vs W4 CMD ({\it right}) of S21 AGB stars (blue dots), P21 YSOs (red stars), and 11 W51 high-mass YSOs (dark blue stars) from \citet{2025Kim}. All high-mass YSOs were classified as YSOs by our Filter 1.}
    \label{HMYSO}
\end{figure*}

{
SPICY is one of the largest Milky Way YSO catalogs, which consists of 117,446 YSO candidates observed by the Spitzer telescope and categorized by the random forest method \citep{2021Kuhn}. The catalog verified these YSO candidates against scientific properties, including IR photometry, spatial clustering, and variability, thereby increasing the reliability of the catalog. Yet, the paper also mentioned that the classification struggled with contaminants other than YSOs, mainly AGB stars \citep{2021Kuhn}. From about 120,000 sources, 27,493 YSOs had all 2MASS, AllWISE, and NEOWISE 20 epochs, which are essential for our Double Filter model.

The P21 catalog does not include high-mass YSOs; therefore, our model may not be directly applicable to the SPICY catalog. While SPICY covers the full mass range of YSOs, the fraction of high-mass YSOs is expected to be very small given the initial mass function. However, as shown in Figure \ref{HMYSO}, the IR colors of high-mass YSOs are significantly redder than those of AGB stars and low-mass YSOs, reducing the likelihood of confusion with AGBs. To validate the model performance for high-mass YSOs, we applied Filter 1 to high-mass YSOs identified in W51 \citep{2025Kim} (Figure \ref{HMYSO}); all objects in this sample were successfully classified as YSOs, confirming that high-mass YSOs can be readily separated from AGBs using only IR magnitudes and colors. Therefore, we conclude that our double-filter model can be reliably applied to the SPICY catalog.

Table \ref{tabE} shows the results of Filter 1 and Filter 2 from the SPICY catalog. Filter 1 classified 26,994 as YSOs and 499 as AGB stars, while Filter 2 returned 241 as YSOs. In total, the Double Filter model reconfirmed 27,235 out of 27,493 sources as YSO candidates, but suspected 258 as tentative AGB star candidates. Our result matches $\sim$ 99 \% of YSOs with \citet{2021Kuhn}, although the classification method, inputs, and dataset they used are different from ours. Detailed information of the 258 AGB candidates is compiled in Table \ref{tab_258_cand_mr}. 
}

\subsection{AGB candidates from the SPICY}

\begin{table}[]
    \centering
    \caption{The results of SPICY YSOs from the model.}
    \label{tabE} %
    \begin{threeparttable}
        \begin{tabular*}{\columnwidth}{@{\extracolsep{\fill}} c *{4}{S[table-format=1.3]}}
            \toprule
            \textbf{}          & \textbf{confirmed YSOs}              & \multicolumn{2}{c}{\textbf{misclassified AGBs}} \\ \hline
            \textbf{Filter 1}  & {26994}                                & \multicolumn{2}{c}{\textcolor{red}{\textbf{499}\tnote{a}}}     \\
            \textbf{Filter 2}  & \textcolor{blue}{\textbf{241}\tnote{b}}       & \multicolumn{2}{c}{\textcolor{blue}{\textbf{258}\tnote{b}}}   \\ \hline
            \textbf{Total}     & \textbf{27235}                       & \multicolumn{2}{c}{\textbf{258}}                \\ \hline
        \end{tabular*} \vspace{5pt}
        \textbf{Notes.}
        \begin{tablenotes}[flushleft]\footnotesize
        \item[a] Misclassified as AGB stars by Filter 1.
        \item[b] Reclassified from [a] by Filter 2. 258 are the final tentative AGB candidates.
        \end{tablenotes}
    \end{threeparttable}%
\end{table}%

\begin{figure}
    \centering
    \includegraphics[width=0.45\textwidth ]{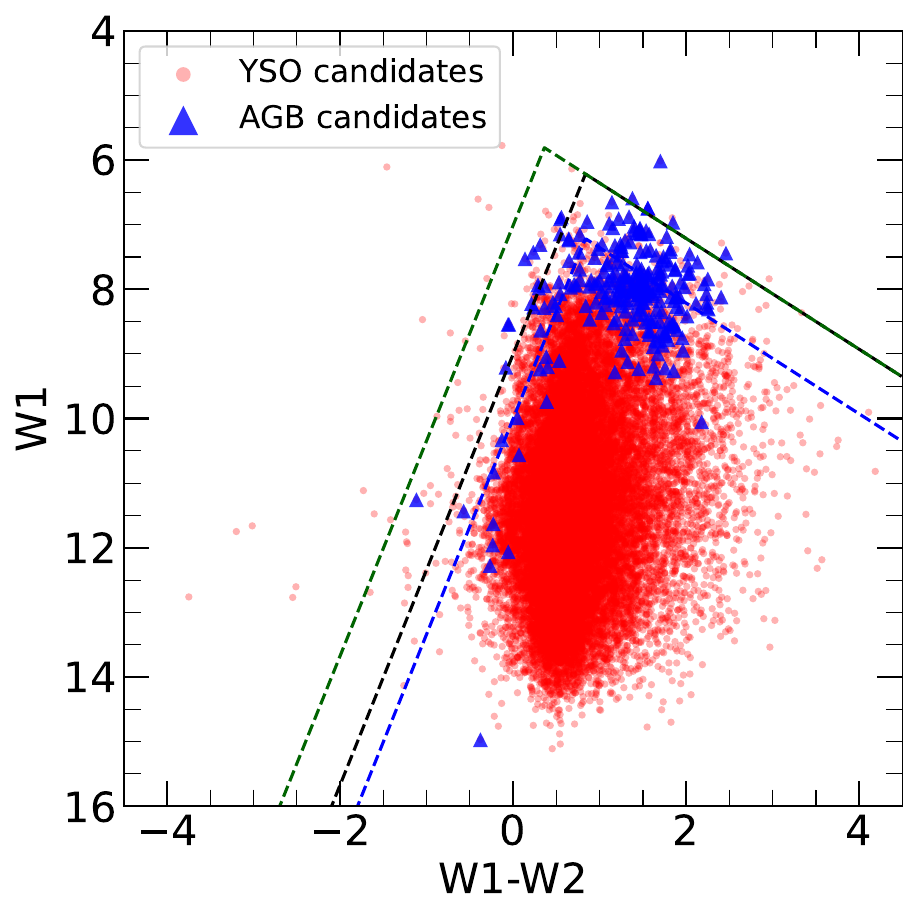}
    \caption{The W1-W2 vs. W1 CMD graph of the SPICY catalog. Red dots are 27,235 YSOs, while blue triangles are 258 AGB candidates from our model. The dashed lines are the same as those in Figure \ref{fig0A}.}
    \label{figG}
\end{figure}

\begin{table*}[]
\centering
\caption{The variable types of SPICY YSOs classified as AGBs in Filter 1, and reclassified YSOs and AGBs in Filter 2.}
\label{tab5}
\begin{tabularx}{\textwidth}{YYYYY}
\hline
\hline
\multicolumn{2}{c}{\bf Variability Type}                                &  {\bf Filter   1 (AGB)}       & {\bf Filter   2 (AGB)} & {\bf Filter   2 (YSO)} \\ \hline
\multicolumn{1}{c}{\multirow{3}{*}{Secular}}    & Periodic  & 235   (47.09 \%) & 219 (84.88 \%)   & 16  (6.64 \%)   \\
\multicolumn{1}{c}{}                            & Curved    & 69   (13.83 \%)  & 10  (3.88 \%)    & 59  (24.48 \%)  \\
\multicolumn{1}{c}{}                            & Linear    & 0   (0.00 \%)    & 0   (0.00 \%)    & 0   (0.00 \%)    \\ \hline
\multicolumn{1}{c}{\multirow{3}{*}{Stochastic}} & Burst     & 0   (0.00 \%)    & 0   (0.00 \%)    & 0   (0.00 \%)    \\
\multicolumn{1}{c}{}                            & Drop      & 0   (0.00 \%)    & 0   (0.00 \%)    & 0   (0.00 \%)    \\
\multicolumn{1}{c}{}                            & Irregular & 26   (5.21 \%)   & 11  (4.26 \%)    & 15   (6.22 \%)    \\ \hline
\multicolumn{2}{c}{Non-variable}                         & 169   (33.87 \%) & 18  (6.98 \%)    & 151  (62.66 \%)  \\ \hline
\multicolumn{2}{c}{\bf Total}                               & {\bf 499}    & {\bf 258}   & {\bf 241}  \\ \hline
\end{tabularx}
\end{table*}

\begin{figure*}
    \centering
    \includegraphics[width=0.45\textwidth ]{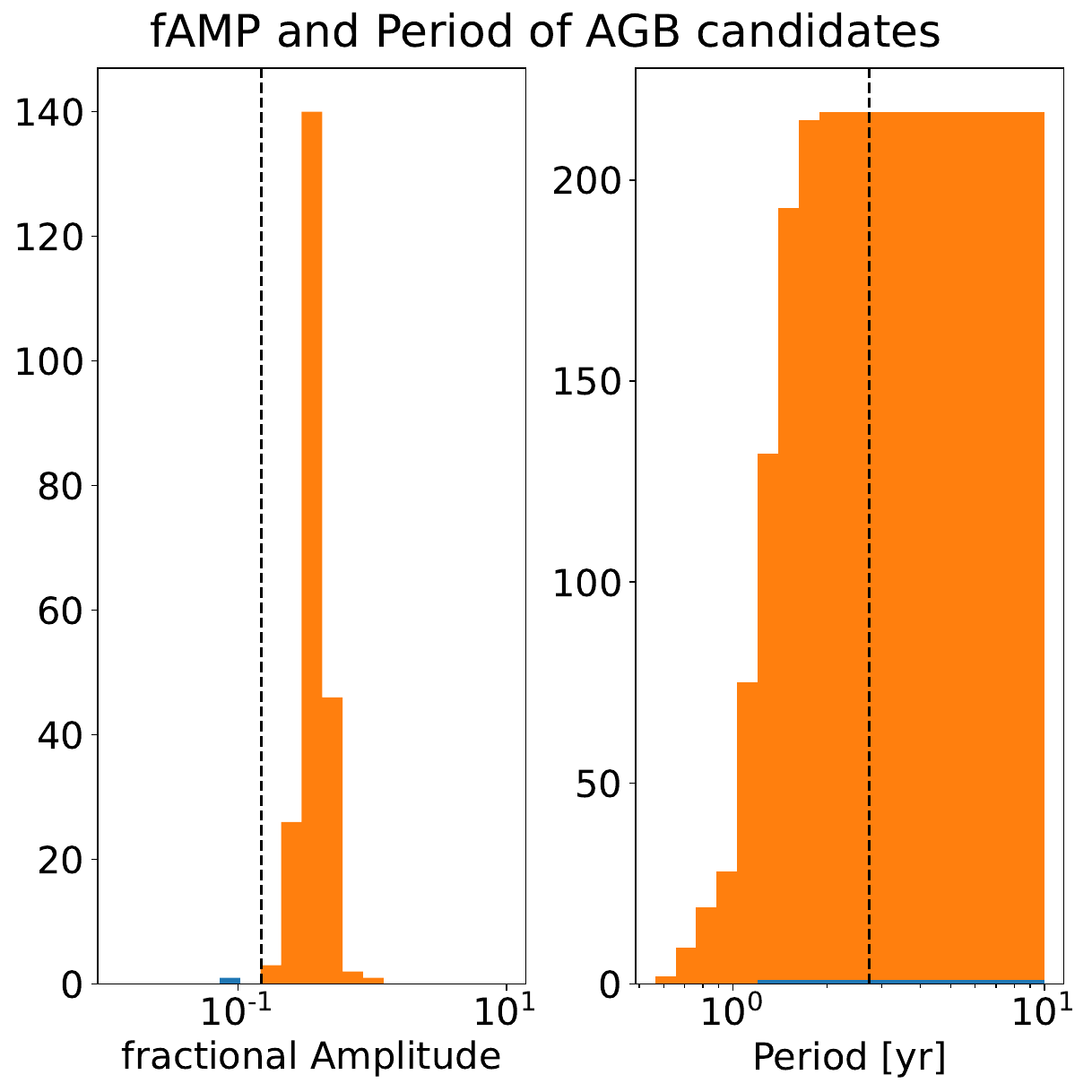}
    \includegraphics[width=0.45\textwidth ]{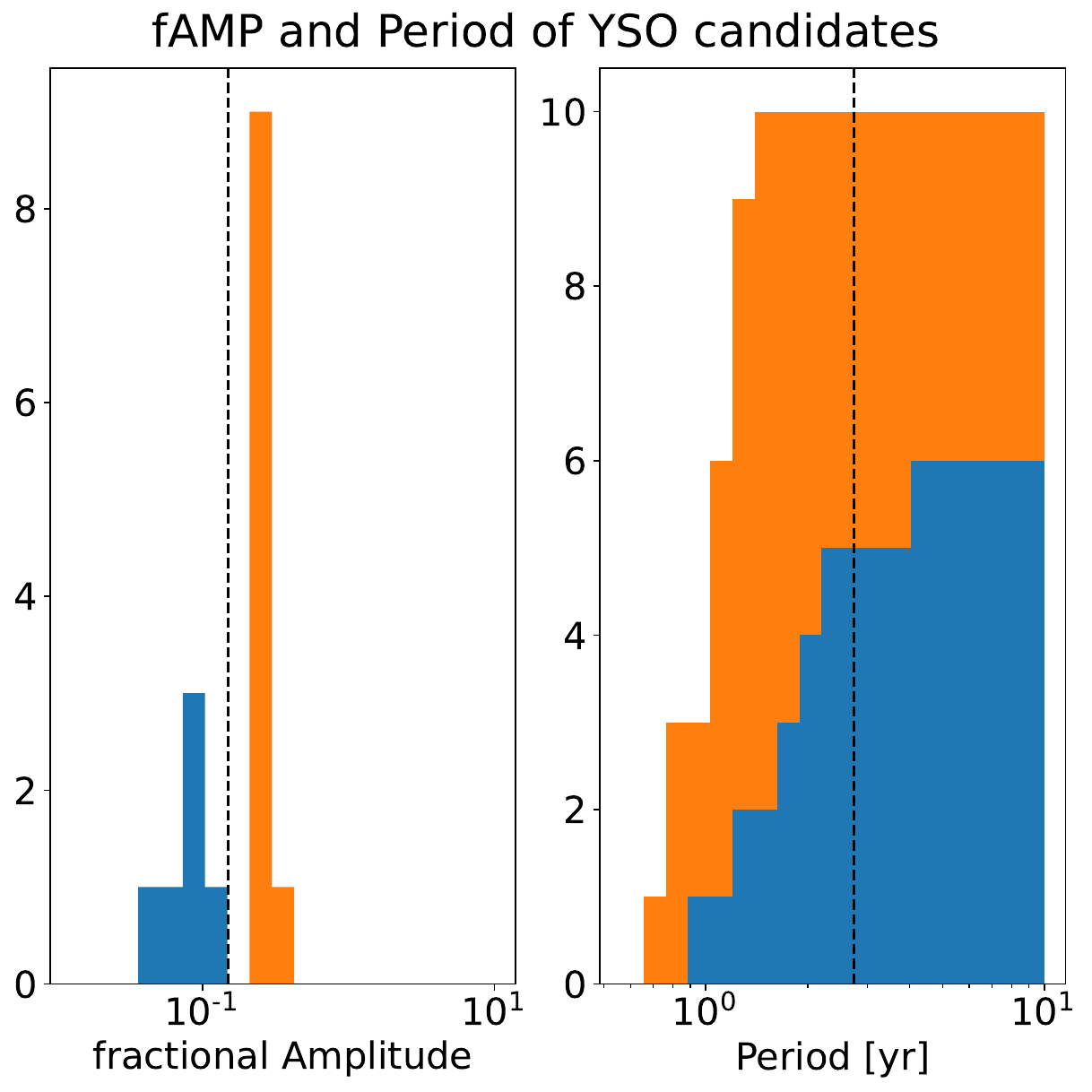}
    \caption{The histograms of periodic light curves for YSO and AGB candidates from SPICY YSOs by Filter 2. The left two histograms are fAMP and the accumulated histogram of periods for 219 periodic AGB candidates, whereas the right two are for the 16 periodic YSO candidates. The black dashed lines represent the criteria of the fAMP with the values of 0.15 according to \citet{2021Park} and period of 1000 days, which is often regarded as the upper limit of the AGB period \citep{2008Whitelock}. The blue histograms are those satisfying fAMP $<$ 0.15, while the orange histograms are the opposite. }
    \label{figH}
\end{figure*}

{
These 258 tentative AGB candidates classified by our model are generally brighter than YSOs, as shown in Figure \ref{figG}. We examined these sources in detail, analyzing their light-curve behavior to interpret the model’s decisions and validate its performance.

Table \ref{tab5} summarizes the light curve characteristics of 499 sources classified as AGB stars by Filter 1, including 258 AGB candidates and 241 reclassified YSO candidates by Filter 2. Given the light curve trend, we contend that Filter 2 captures periodicity in the data: 219 of 258 AGB candidates (84.88 \%) showed periodic light curves, while only 16 of 241 YSO candidates (6.64 \%) did so.

An in-depth analysis of the light curves of these candidates supports them as plausible AGB stars. As mentioned in \ref{sec2-2-2}, AGB stars are high-amplitude (fAMP $>$ 0.15) long-period variables (LPVs) with periods ranging from 100 days to 1000 days \citep{2008Whitelock, 2018Hofner}. In contrast, YSOs show irregular or stochastic light curves, leaving only 1 $\sim$ 2 \% of YSOs as periodic \citep{2021Lee} with fAMP lower than 0.15 \citep{2021Park}.

Among 219 periodic AGB candidates, 218 sources satisfied with fAMP over 0.15 and period under 1000 days (2.74 years), as shown in the first and second histograms in Figure \ref{figH}. Therefore, these AGB candidates are likely to be true AGB stars. For the 16 periodic YSO candidates, only 6 satisfied the fAMP criterion, with values below 0.15. As YSOs can exhibit periodic light curves, but periods of about a few hundred days are extremely rare \citep[e.g., EC 53,][]{2021LeeY}, the remaining 10 YSO candidates are likely due to model errors. Consequently, we estimate that the model is worthwhile in identifying hidden AGB contaminants that the photometry method alone cannot distinguish; however, it is questionable for reclassified periodic YSOs, which require additional confirmation.

One possible method to confirm its classification is to observe SiO masers or carbon-bearing molecules. AGB stars are divided into two broad cases: oxygen-rich AGB stars (OAGBs) and carbon-rich AGB stars (CAGBs). OAGBs emit SiO or OH masers, while YSOs generally do not \citep{1998Nyman, 2021Lee}. Thus, if we detect these masers among the OAGB candidates, we can confirm them as true OAGBs. For CAGBs, NIR features of C$_2$ and C$_2$H$_2$ can be used to distinguish CAGBs from YSOs \citep{2006Matsuura, 2009Wright}. Some of the AGB candidates identified in this work have already been confirmed through spectroscopic observations, and the results will be published in separate papers.
}

\section{Discussion} \label{sec6}

\begin{figure*}
    \centering
    \includegraphics[width=0.7\textwidth ]{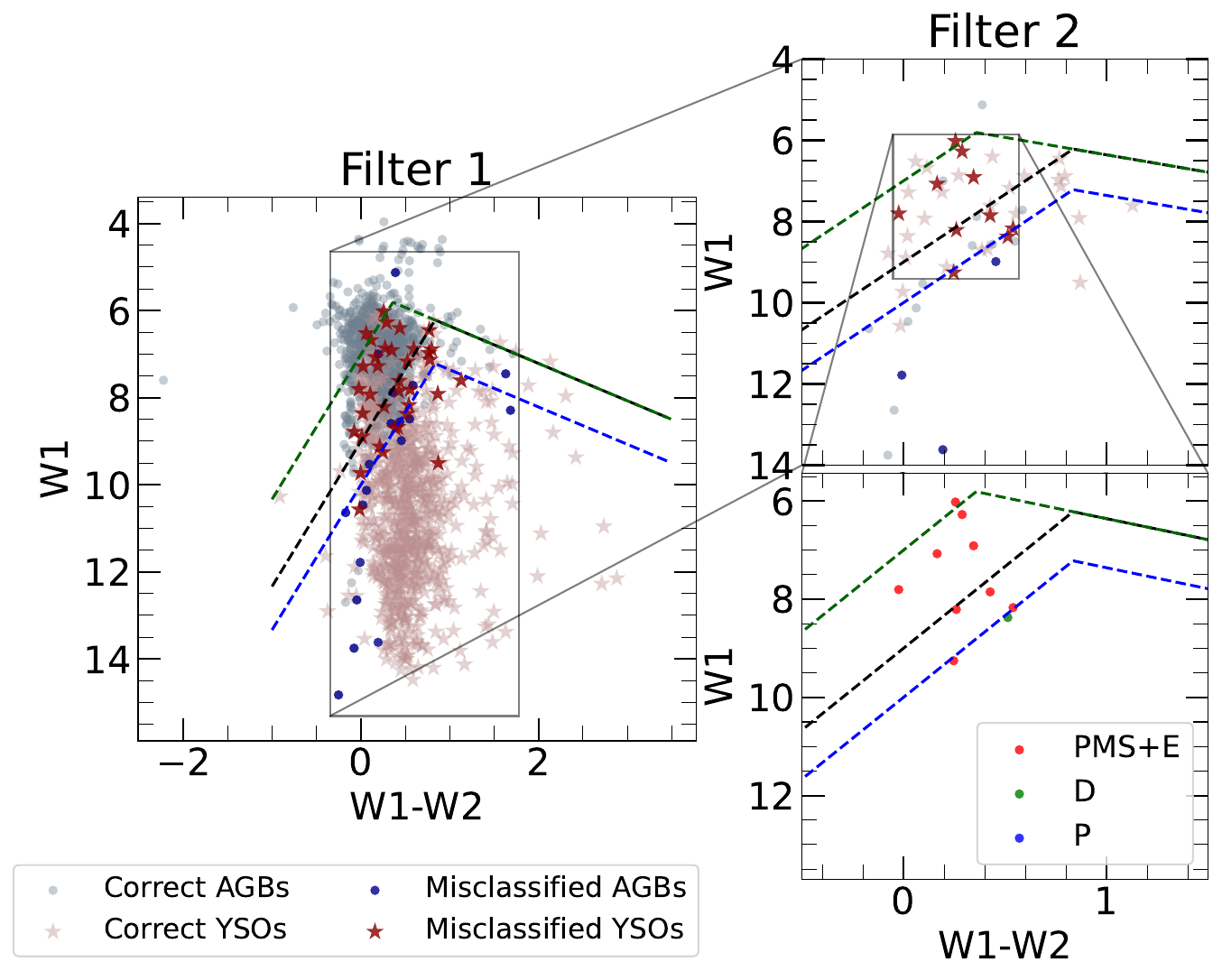}
    \caption{\textit{Left and top right}: The CMD (W1-W2 vs. W1) graphs of YSOs and AGB stars classified by Filter 1 (\textit{left}) and Filter 2 model (\textit{top right}). Light red stars and blue dots are correctly classified as YSOs and AGB stars, while dark red stars and dark blue dots are the misclassified YSOs and AGB stars from each Filter of the model.
    {\it Bottom right}: The YSO sub-classes of 10 misclassified YSOs with Filter 2 from the test data of P21 YSOs. Red, green, and blue dots represent pre-main-sequence YSOs and/or Evolved stars (PMS+E), Class II disks (D), and Class 0/I protostars (P), respectively. 9 of 10 are revealed as PMS+E and remaining 1 as D. None of P were misclassified. The dashed lines are those in Figure \ref{fig0A}}
    \label{figC}
\end{figure*}

\begin{table*}[]
\centering
\caption{A detailed information about the final misclassified 10 YSOs and 3 AGB stars of the test data by the model.}
\label{tab_dis}
\begin{threeparttable}
\begin{tabular*}{\textwidth}{@{\extracolsep{\fill}} l *{5}{c} l} %{S[table-format=1.0]}}
\toprule
 \multicolumn{1}{c}{\multirow{2}{*}{\textbf{ID\tnote{a}}}} & \multirow{2}{*}{\textbf{RA (ICRS)}} & \multirow{2}{*}{\textbf{Dec (ICRS)}} & \multicolumn{3}{c}{\textbf{Results}} & \multicolumn{1}{c}{\multirow{2}{*}{\textbf{Reference}}} \\ \cline{4-6}
               &             &                &  \textbf{Subclass\tnote{b}} & \textbf{SIMBAD} & \textbf{Variability\tnote{c}}  \\ \hline
\textbf{D79}   & 18:23:09.87 & -03:10:35.41   & PMS+E & LPV           && \footnotesize{\citet{2023GaiaAGB}} \\
\textbf{D204}  & 18:28:07.44 & -02:58:35.53   & PMS+E & AGB candidate &Linear& \footnotesize{\citet{2015Dunham}} \\
\textbf{D377}  & 18:29:23.26 & -02:59:06.21   & PMS+E & LPV && \footnotesize{\citet{2023GaiaAGB}} \\
\textbf{D1256} & 18:39:05.46 & +00:06:05.32   & PMS+E & LPV           && \footnotesize{\citet{2023GaiaAGB}} \\
\textbf{D1669} & 12:37:54.32 & -80:21:30.28   & PMS+E & AGB candidate &Periodic& \footnotesize{\citet{2023GaiaAGB}} \\
\textbf{D1877} & 16:07:54.09 & -39:20:46.27   & PMS+E & LPV           && \footnotesize{\citet{2023GaiaAGB}} \\
\textbf{D2016} & 16:23:49.04 & -40:26:17.59   & PMS+E & AGB           && \footnotesize{\citet{2012Romero}} \\
\textbf{D2050} & 12:34:51.22 & -70:30:51.17   & D     & LPV           &Curved& \footnotesize{\citet{2023GaiaAGB}} \\
\textbf{D2342} & 16:44:29.33 & -24:15:55.59   & PMS+E & AGB           && \footnotesize{\citet{2023GaiaAGB}} \\
\textbf{D2770} & 18:28:50.40 & -00:12:55.36   & PMS+E & LPV           &Irregular& \footnotesize{\citet{2023GaiaAGB}} \\
\textbf{6719\tnote{d}}  & 17:42:03.97 & -24:59:21.43   & O-rich AGB & Mira         &Irregular& \footnotesize{\citet{2023Iwanek}}\\
\textbf{16663} & 15:19:05.98 & +50:07:02.86   & C-rich AGB & Multiple star &&  \footnotesize{\citet{2021Whitehouse}} \\
\textbf{16047} & 10:40:06.37 & +35:48:02.38   & C-rich AGB & Chemically Peculiar star && \footnotesize{\citet{2021Whitehouse}} \\ \hline
\end{tabular*} \vspace{5pt}
    \textbf{Notes.}
    \begin{tablenotes}[flushleft]\footnotesize
        \item[a] Sources' ID starting with a letter D are from \citet{2021Park}, and starting with numbers are from \citet{2021Suh}.
        \item[b] Subclasses of PMS+E and D are from \citet{2021Park}, and O-rich AGB and C-rich AGB are from \citet{2021Suh}.
        \item[c] Categorized based on Lomb-Scargle Periodogram. Blanks are `Non-variables'.
        \item[d] Also included in SPICY \citep{2021Kuhn} catalog.
        \end{tablenotes}
\end{threeparttable}
\end{table*}

\subsection{Model errors from the test data}
{
The Double Filter model achieved performance exceeding 99\% on the test dataset (see Table \ref{tab4}). Despite its high accuracy, ML is often referred to as a `black box' due to its non-linear process and limited interpretability. Therefore, understanding the precise rationale behind individual ML decisions is challenging. Yet, some post-analysis with scientific reasoning enabled us to infer the model's decision more restrictively.

One approach is to examine their distribution on W1-W2 vs. W1 CMD. Figure \ref{figC} shows the W1-W2 vs. W1 CMD of the test data with our model. The misclassified YSOs and AGB stars (dark red stars and dark blue dots) in the left plot of Figure \ref{figC} confirm that Filter 1 follows the empirical classification by other studies, such as brighter sources are regarded as AGB stars while dimmer sources are regarded as YSOs \citep{2014K&L, 2024Suh}. This implies that Filter 1 catches the general IR photometric features that other studies have drawn \citep{2009Gutermuth, 2014K&L}. However, it fails to clearly classify sources within the overlapping region, as they show no significant differences in color or magnitude.

Filter 2, on the other hand, returned more than half of these misclassified sources to their original categories, leaving only 10 YSOs and 3 AGB stars misclassified. 
Cross-matching with other studies or catalogs can help reveal these confusing sources through additional analysis. The misclassified YSOs and AGB stars identified by Filter 2 are listed in the {\it Set of Identifications, Measurements, and Bibliography for Astronomical Data} (SIMBAD), which provides information about the sources, along with relevant research and references. Table \ref{tab_dis} shows the details of 10 misclassified YSOs and 3 AGB stars. Subclasses are originated from \citet{2021Park} and \citet{2021Suh}, and variability is identical to the method used in Table \ref{tab2-1} and Table \ref{tab5}. The SIMBAD class states that all 10 misclassified YSOs are considered as AGB stars. This result indicates that Filter 2 functioned as expected. 3 misclassified AGB stars, which are regarded as evolved stars according to SIMBAD, draw a tentative conclusion that their misclassification is due to the model's limitations.  

The two approaches support the model's performance, despite their limited interpretation. The CMD method assesses the suitability of Filter 1, while SIMBAD or comparisons with other studies assess Filter 2. Therefore, we estimate that the double filter model is adequate for YSO-AGB classification.

}

\subsection{Model caveats}
{
Although our model demonstrates strong performance,  two major caveats should be considered. One notable concern is the purity of the original YSO or AGB catalogs. Since our model is designed for binary classification, any sources in the catalogs that are neither YSOs nor AGB stars will inevitably be classified as either. As a result, any sources, such as main sequence (MS) stars, post-AGB (PAGB) stars, planetary nebulae (PNe), or extragalactic sources included in the catalogs cannot be filtered out. It is, therefore, crucial to preemptively identify and exclude such unwanted sources to ensure classification accuracy.  

Fortunately, the catalogs used for training, validation, and application in this study were carefully curated. These catalogs primarily consist of sources observed mainly in the Galactic midplane and star-forming regions \citep{2021Suh, 2021Park, 2021Kuhn}, where IR-bright extragalactic sources are typically fainter than YSOs and are likely to be obscured by dust extinction \citep{2016Marton, 2021Kuhn}. Reddened MS stars, on the other hand, can be effectively distinguished using NIR CMD and CCD criteria \citep{2019Akras} due to the non-existence of IR excess \citep{2004Allen}. PNe can be preemptively distinguished from YSOs using IR CCDs \citep{2024Suh}.

However, the SPICY catalog may include PAGB stars \citep{2021Kuhn}. Given that PAGB stars have lifetimes roughly two orders of magnitude shorter than those of AGB stars \citep[][]{2005Herwig, 2016Miguel}, their contribution is expected to be small. Nevertheless, PAGB stars that lack pulsation and show extremely red infrared colors may still be misclassified as YSOs. 

The other significant concern is `data evaporation', which occurs during data preprocessing. The original catalogs of P21 and S21 contain 5391 YSOs and 8877 AGB stars, respectively. However, these numbers decrease to 4023 YSOs and 5730 AGB stars when cross-matched with 2MASS and NEOWISE data. It is estimated that about 26 \% to 35 \% of data loss took place, potentially increasing the concerns about data bias and insufficient training of ML models. Although our model's performance remained robust through the implementation of data augmentation and dropout techniques \citep{2014Dropout, 2017DataAugmentation}, continued attention to this issue will be crucial for future applications.

Unfortunately, the external catalogs we applied to our model experienced a critical data loss. The validation of AGBs and SPICY YSO catalogs suffered significant data loss, up to 90 \%. The losses are mainly due to the requirement of 20 epochs of NEOWISE observation. The data-loss issue is well recognized in the machine learning field as the `missing values imputation' problem \citep{2021missingvalue, 2024Ye}. However, rather than applying imputation methods, we excluded all sources that lacked NEOWISE data or were only partially covered to maintain model integrity. 
}
 
\section{Summary} \label{sec7}
{
We developed an ML model to refine the SPICY catalog, addressing the limitations inherent in using only IR photometric magnitudes and colors for classification. Our approach leverages NEOWISE time-series data, which provides biannual observations in two IR bands (W1 and W2) from 2013 to 2023. Integrating both single-epoch photometric data and NEOWISE light curve, we constructed a two-stage classification model, named the `Double Filter model'.

Filter 1 of the model classifies sources using IR photometry (2MASS J, H, K bands and AllWISE W1, W2, W3, and W4 bands) with three ML methods. Misclassified data from Filter 1 are then re-evaluated in Filter 2, using 20 epochs of NEOWISE W1 and W2 fluxes, their errors, W1-W2 color, $\Delta$W1/$\sigma$, $\Delta$W2/$\sigma$, and additional time-domain features such as period and fractional amplitude, calculated through the Lomb-Scargle periodogram. This double-filter approach improved classification reliability by leveraging both static and time-variable properties. 

The Double Filter model achieved a classification accuracy of 99.3 \% for YSOs and AGB stars on the test dataset. To further validate the model, we applied it to confirmed Taurus YSOs and spectroscopically studied AGB stars. The model demonstrated its classification, showing all Taurus YSOs as YSOs and 88.2 \% of AGB stars as AGB stars. It was also able to successfully classify AGB stars in the LMC, which can be challenging because they are dimmer and redder than nearby AGB stars.

We applied our model to the SPICY YSO catalog \citep{2021Kuhn}. From 27,493 YSO candidates with the appropriate data to use the model, we found 258 potential AGB stars. A detailed analysis of these AGB interlopers reveals that they are predominantly bright sources exhibiting periodic light curves. This highlights that these candidates may indeed be actual AGB stars. 
}

\begin{acknowledgments}

This work was supported by the NRF grant funded by the Korean government (MSIT) (grant numbers 2021R1A2C1011718 and RS-2024-00416859). CCP was supported by the NRF grant funded by the Korean government (MEST) (grant numbers. 2019R1A6A1A10073437). H. Jheonn gives special thanks to Yeonjoon Jung, a Ph.D. student in the Department of Computer Science and Engineering at Seoul National University, for providing insights and advice on various machine learning and deep learning methods. We also appreciate the insightful comments of Doug Johnstone and Gregory Herczeg. Finally, the authors are grateful to the anonymous referee for valuable comments and suggestions, which significantly improved this paper.

\end{acknowledgments}

\bibliographystyle{aasjournalv7}
\bibliography{main}{}

\begin{thebibliography}{}
\expandafter\ifx\csname natexlab\endcsname\relax\def\natexlab#1{#1}\fi
\providecommand{\url}[1]{\href{#1}{#1}}
\providecommand{\dodoi}[1]{doi:~\href{http://doi.org/#1}{\nolinkurl{#1}}}
\providecommand{\doeprint}[1]{\href{http://ascl.net/#1}{\nolinkurl{http://ascl.net/#1}}}
\providecommand{\doarXiv}[1]{\href{https://arxiv.org/abs/#1}{\nolinkurl{https://arxiv.org/abs/#1}}}

\bibitem[{S. {Akras} {et~al.}(2019){Akras}, {Leal-Ferreira}, {Guzman-Ramirez}, \& {Ramos-Larios}}]{2019Akras}
{Akras}, S., {Leal-Ferreira}, M.~L., {Guzman-Ramirez}, L., \& {Ramos-Larios}, G. 2019, \bibinfo{title}{{A machine learning approach for identification and classification of symbiotic stars using 2MASS and WISE},} \mnras, 483, 5077, \dodoi{10.1093/mnras/sty3359}

\bibitem[{L.~E. {Allen} {et~al.}(2004){Allen}, {Calvet}, {D'Alessio}, {Merin}, {Hartmann}, {Megeath}, {Gutermuth}, {Muzerolle}, {Pipher}, {Myers}, \& {Fazio}}]{2004Allen}
{Allen}, L.~E., {Calvet}, N., {D'Alessio}, P., {et~al.} 2004, \bibinfo{title}{{Infrared Array Camera (IRAC) Colors of Young Stellar Objects},} \apjs, 154, 363, \dodoi{10.1086/422715}

\bibitem[{P. {Andre} {et~al.}(1993){Andre}, {Ward-Thompson}, \& {Barsony}}]{1993Andre}
{Andre}, P., {Ward-Thompson}, D., \& {Barsony}, M. 1993, \bibinfo{title}{{Submillimeter Continuum Observations of rho Ophiuchi A: The Candidate Protostar VLA 1623 and Prestellar Clumps},} \apj, 406, 122, \dodoi{10.1086/172425}

\bibitem[{J. {Bally}(2016){Bally}}]{2016Bally}
{Bally}, J. 2016, \bibinfo{title}{{Protostellar Outflows},} \araa, 54, 491, \dodoi{10.1146/annurev-astro-081915-023341}

\bibitem[{L. Breiman(2001)Breiman}]{2001RF}
Breiman, L. 2001, \bibinfo{title}{Random Forests,} Machine Learning, 45, 5, \dodoi{10.1023/A:1010933404324}

\bibitem[{Z. {Chen} {et~al.}(2025){Chen}, {Johnstone}, {Contreras Pe{\~n}a}, {Lee}, {Liu}, {Herczeg}, {Mairs}, {Park}, {Kim}, {Kim}, {Qiu}, {Wang}, {Zhang}, {Reiter}, \& {JCMT Transient Team}}]{2025Chen}
{Chen}, Z., {Johnstone}, D., {Contreras Pe{\~n}a}, C., {et~al.} 2025, \bibinfo{title}{{Submillimeter and Mid-infrared Variability of Young Stellar Objects in the M17 H II Region},} \aj, 170, 125, \dodoi{10.3847/1538-3881/ade988}

\bibitem[{C. {Contreras Pe{\~n}a} {et~al.}(2019){Contreras Pe{\~n}a}, {Naylor}, \& {Morrell}}]{2019Carlos}
{Contreras Pe{\~n}a}, C., {Naylor}, T., \& {Morrell}, S. 2019, \bibinfo{title}{{Determining the recurrence time-scale of long-lasting YSO outbursts},} \mnras, 486, 4590, \dodoi{10.1093/mnras/stz1019}

\bibitem[{C. Cortes \& V. Vapnik(1995)Cortes \& Vapnik}]{1995SVM}
Cortes, C., \& Vapnik, V. 1995, \bibinfo{title}{Support-vector networks,} Machine Learning, 20, 273, \dodoi{10.1007/BF00994018}

\bibitem[{T.~G. Dietterich(2000)Dietterich}]{2000Ensemble}
Dietterich, T.~G. 2000, in Multiple Classifier Systems (Berlin, Heidelberg: Springer Berlin Heidelberg), 1--15

\bibitem[{M.~M. {Dunham} {et~al.}(2015){Dunham}, {Allen}, {Evans}, {Broekhoven-Fiene}, {Cieza}, {Di Francesco}, {Gutermuth}, {Harvey}, {Hatchell}, {Heiderman}, {Huard}, {Johnstone}, {Kirk}, {Matthews}, {Miller}, {Peterson}, \& {Young}}]{2015Dunham}
{Dunham}, M.~M., {Allen}, L.~E., {Evans}, Neal~J., I., {et~al.} 2015, \bibinfo{title}{{Young Stellar Objects in the Gould Belt},} \apjs, 220, 11, \dodoi{10.1088/0067-0049/220/1/11}

\bibitem[{S. {Dutta} {et~al.}(2022){Dutta}, {Yang}, {Bernardis}, {Dobriban}, \& {Lee}}]{2022Dutta}
{Dutta}, S., {Yang}, Y., {Bernardis}, E., {Dobriban}, E., \& {Lee}, I. 2022, \bibinfo{title}{{Memory Classifiers: Two-stage Classification for Robustness in Machine Learning},} arXiv e-prints, arXiv:2206.05323, \dodoi{10.48550/arXiv.2206.05323}

\bibitem[{T. Emmanuel {et~al.}(2021)Emmanuel, Maupong, Mpoeleng, Semong, Mphago, \& Tabona}]{2021missingvalue}
Emmanuel, T., Maupong, T., Mpoeleng, D., {et~al.} 2021, \bibinfo{title}{A survey on missing data in machine learning,} Journal of Big Data, 8, 140, \dodoi{10.1186/s40537-021-00516-9}

\bibitem[{T.~L. {Esplin} \& K.~L. {Luhman}(2019){Esplin} \& {Luhman}}]{2019Esplin}
{Esplin}, T.~L., \& {Luhman}, K.~L. 2019, \bibinfo{title}{{A Survey for New Members of Taurus from Stellar to Planetary Masses},} \aj, 158, 54, \dodoi{10.3847/1538-3881/ab2594}

\bibitem[{A. {Frank} {et~al.}(2014){Frank}, {Ray}, {Cabrit}, {Hartigan}, {Arce}, {Bacciotti}, {Bally}, {Benisty}, {Eisl{\"o}ffel}, {G{\"u}del}, {Lebedev}, {Nisini}, \& {Raga}}]{2014Frank}
{Frank}, A., {Ray}, T.~P., {Cabrit}, S., {et~al.} 2014, in Protostars and Planets VI, ed. H.~{Beuther}, R.~S. {Klessen}, C.~P. {Dullemond}, \& T.~{Henning}, 451--474, \dodoi{10.2458/azu_uapress_9780816531240-ch020}

\bibitem[{P.~A.~B. {Galli} {et~al.}(2018){Galli}, {Loinard}, {Ortiz-L{\'e}on}, {Kounkel}, {Dzib}, {Mioduszewski}, {Rodr{\'\i}guez}, {Hartmann}, {Teixeira}, {Torres}, {Rivera}, {Boden}, {Evans}, {Brice{\~n}o}, {Tobin}, \& {Heyer}}]{2018Galli}
{Galli}, P. A.~B., {Loinard}, L., {Ortiz-L{\'e}on}, G.~N., {et~al.} 2018, \bibinfo{title}{{The Gould's Belt Distances Survey (GOBELINS). IV. Distance, Depth, and Kinematics of the Taurus Star-forming Region},} \apj, 859, 33, \dodoi{10.3847/1538-4357/aabf91}

\bibitem[{M.~A. {Ganaie} {et~al.}(2021){Ganaie}, {Hu}, {Malik}, {Tanveer}, \& {Suganthan}}]{2021Ma_Ensemble_Review}
{Ganaie}, M.~A., {Hu}, M., {Malik}, A.~K., {Tanveer}, M., \& {Suganthan}, P.~N. 2021, \bibinfo{title}{{Ensemble deep learning: A review},} arXiv e-prints, arXiv:2104.02395, \dodoi{10.48550/arXiv.2104.02395}

\bibitem[{M. {Grandini} {et~al.}(2020){Grandini}, {Bagli}, \& {Visani}}]{2020Grandini}
{Grandini}, M., {Bagli}, E., \& {Visani}, G. 2020, \bibinfo{title}{{Metrics for Multi-Class Classification: an Overview},} arXiv e-prints, arXiv:2008.05756, \dodoi{10.48550/arXiv.2008.05756}

\bibitem[{T.~P. {Greene} {et~al.}(1994){Greene}, {Wilking}, {Andre}, {Young}, \& {Lada}}]{1994Greene}
{Greene}, T.~P., {Wilking}, B.~A., {Andre}, P., {Young}, E.~T., \& {Lada}, C.~J. 1994, \bibinfo{title}{{Further Mid-Infrared Study of the rho Ophiuchi Cloud Young Stellar Population: Luminosities and Masses of Pre--Main-Sequence Stars},} \apj, 434, 614, \dodoi{10.1086/174763}

\bibitem[{R. {Guandalini} {et~al.}(2006){Guandalini}, {Busso}, {Ciprini}, {Silvestro}, \& {Persi}}]{2006Guandalini}
{Guandalini}, R., {Busso}, M., {Ciprini}, S., {Silvestro}, G., \& {Persi}, P. 2006, \bibinfo{title}{{Infrared photometry and evolution of mass-losing AGB stars. I. Carbon stars revisited},} \aap, 445, 1069, \dodoi{10.1051/0004-6361:20053208}

\bibitem[{Z. {Guo} {et~al.}(2021){Guo}, {Lucas}, {Contreras Pe{\~n}a}, {Smith}, {Morris}, {Kurtev}, {Borissova}, {Alonso-Garc{\'\i}a}, {Minniti}, {Chen{\'e}}, {Kumar}, {Caratti o Garatti}, {Froebrich}, \& {Stimson}}]{2021Guo}
{Guo}, Z., {Lucas}, P.~W., {Contreras Pe{\~n}a}, C., {et~al.} 2021, \bibinfo{title}{{Analysis of physical processes in eruptive YSOs with near-infrared spectra and multiwavelength light curves},} \mnras, 504, 830, \dodoi{10.1093/mnras/stab882}

\bibitem[{R.~A. {Gutermuth} {et~al.}(2009){Gutermuth}, {Megeath}, {Myers}, {Allen}, {Pipher}, \& {Fazio}}]{2009Gutermuth}
{Gutermuth}, R.~A., {Megeath}, S.~T., {Myers}, P.~C., {et~al.} 2009, \bibinfo{title}{{A Spitzer Survey of Young Stellar Clusters Within One Kiloparsec of the Sun: Cluster Core Extraction and Basic Structural Analysis},} \apjs, 184, 18, \dodoi{10.1088/0067-0049/184/1/18}

\bibitem[{H.~J. {Habing}(1996){Habing}}]{1996Habing}
{Habing}, H.~J. 1996, \bibinfo{title}{{Circumstellar envelopes and Asymptotic Giant Branch stars},} \aapr, 7, 97, \dodoi{10.1007/PL00013287}

\bibitem[{L. {Hartmann} {et~al.}(2016){Hartmann}, {Herczeg}, \& {Calvet}}]{2016Hartmann}
{Hartmann}, L., {Herczeg}, G., \& {Calvet}, N. 2016, \bibinfo{title}{{Accretion onto Pre-Main-Sequence Stars},} \araa, 54, 135, \dodoi{10.1146/annurev-astro-081915-023347}

\bibitem[{K. {He} {et~al.}(2015){He}, {Zhang}, {Ren}, \& {Sun}}]{2015He}
{He}, K., {Zhang}, X., {Ren}, S., \& {Sun}, J. 2015, \bibinfo{title}{{Deep Residual Learning for Image Recognition},} arXiv e-prints, arXiv:1512.03385, \dodoi{10.48550/arXiv.1512.03385}

\bibitem[{F. {Herwig}(2005){Herwig}}]{2005Herwig}
{Herwig}, F. 2005, \bibinfo{title}{{Evolution of Asymptotic Giant Branch Stars},} \araa, 43, 435, \dodoi{10.1146/annurev.astro.43.072103.150600}

\bibitem[{G.~E. {Hinton} {et~al.}(2012){Hinton}, {Srivastava}, {Krizhevsky}, {Sutskever}, \& {Salakhutdinov}}]{2012Hinton}
{Hinton}, G.~E., {Srivastava}, N., {Krizhevsky}, A., {Sutskever}, I., \& {Salakhutdinov}, R.~R. 2012, \bibinfo{title}{{Improving neural networks by preventing co-adaptation of feature detectors},} arXiv e-prints, arXiv:1207.0580, \dodoi{10.48550/arXiv.1207.0580}

\bibitem[{S. {H{\"o}fner} \& H. {Olofsson}(2018){H{\"o}fner} \& {Olofsson}}]{2018Hofner}
{H{\"o}fner}, S., \& {Olofsson}, H. 2018, \bibinfo{title}{{Mass loss of stars on the asymptotic giant branch. Mechanisms, models and measurements},} \aapr, 26, 1, \dodoi{10.1007/s00159-017-0106-5}

\bibitem[{P. {Iwanek} {et~al.}(2023){Iwanek}, {Poleski}, {Koz{\l}owski}, {Soszy{\'n}ski}, {Pietrukowicz}, {Ban}, {Skowron}, {Mr{\'o}z}, {Wrona}, {Udalski}, {Szyma{\'n}ski}, {Skowron}, {Ulaczyk}, {Gromadzki}, {Rybicki}, \& {Ratajczak}}]{2023Iwanek}
{Iwanek}, P., {Poleski}, R., {Koz{\l}owski}, S., {et~al.} 2023, \bibinfo{title}{{A Three-dimensional Map of the Milky Way Using 66,000 Mira Variable Stars},} \apjs, 264, 20, \dodoi{10.3847/1538-4365/acad7a}

\bibitem[{M.-R. {Kim} {et~al.}(2025){Kim}, {Lee}, {Contreras Pe{\~n}a}, {Herczeg}, {Johnstone}, \& {Kang}}]{2025Kim}
{Kim}, M.-R., {Lee}, J.-E., {Contreras Pe{\~n}a}, C., {et~al.} 2025, \bibinfo{title}{{YSO Variability in the W51 Star-Forming Region},} Journal of Korean Astronomical Society, 58, 231, \dodoi{10.5303/JKAS.2025.58.2.231}

\bibitem[{X.~P. {Koenig} \& D.~T. {Leisawitz}(2014){Koenig} \& {Leisawitz}}]{2014K&L}
{Koenig}, X.~P., \& {Leisawitz}, D.~T. 2014, \bibinfo{title}{{A Classification Scheme for Young Stellar Objects Using the Wide-field Infrared Survey Explorer AllWISE Catalog: Revealing Low-density Star Formation in the Outer Galaxy},} \apj, 791, 131, \dodoi{10.1088/0004-637X/791/2/131}

\bibitem[{X.~P. {Koenig} {et~al.}(2012){Koenig}, {Leisawitz}, {Benford}, {Rebull}, {Padgett}, \& {Assef}}]{2012Koenig}
{Koenig}, X.~P., {Leisawitz}, D.~T., {Benford}, D.~J., {et~al.} 2012, \bibinfo{title}{{Wide-field Infrared Survey Explorer Observations of the Evolution of Massive Star-forming Regions},} \apj, 744, 130, \dodoi{10.1088/0004-637X/744/2/130}

\bibitem[{M.~A. {Kuhn} {et~al.}(2021){Kuhn}, {de Souza}, {Krone-Martins}, {Castro-Ginard}, {Ishida}, {Povich}, {Hillenbrand}, \& {COIN Collaboration}}]{2021Kuhn}
{Kuhn}, M.~A., {de Souza}, R.~S., {Krone-Martins}, A., {et~al.} 2021, \bibinfo{title}{{SPICY: The Spitzer/IRAC Candidate YSO Catalog for the Inner Galactic Midplane},} \apjs, 254, 33, \dodoi{10.3847/1538-4365/abe465}

\bibitem[{C.~J. {Lada}(1987){Lada}}]{1987Lada}
{Lada}, C.~J. 1987, in Star Forming Regions, ed. M.~{Peimbert} \& J.~{Jugaku}, Vol. 115, 1

\bibitem[{K. {Lakshmipathaiah} {et~al.}(2023){Lakshmipathaiah}, {Vig}, {Ashby}, {Hora}, {Kang}, \& {Gorthi}}]{2023Laks}
{Lakshmipathaiah}, K., {Vig}, S., {Ashby}, M. L.~N., {et~al.} 2023, \bibinfo{title}{{Probabilistic classification of infrared-selected targets for SPHEREx mission: in search of young stellar objects},} \mnras, 526, 1923, \dodoi{10.1093/mnras/stad2782}

\bibitem[{T. {Lebzelter} {et~al.}(2023){Lebzelter}, {Mowlavi}, {Lecoeur-Taibi}, {Trabucchi}, {Audard}, {Garc{\'\i}a-Lario}, {Gavras}, {Holl}, {Jevardat de Fombelle}, {Nienartowicz}, {Rimoldini}, \& {Eyer}}]{2023GaiaAGB}
{Lebzelter}, T., {Mowlavi}, N., {Lecoeur-Taibi}, I., {et~al.} 2023, \bibinfo{title}{{Gaia Data Release 3. The second Gaia catalogue of long-period variable candidates},} \aap, 674, A15, \dodoi{10.1051/0004-6361/202244241}

\bibitem[{J.-E. {Lee} {et~al.}(2021){Lee}, {Lee}, {Lee}, {Suh}, {Cho}, {Byun}, {Park}, {Herczeg}, {Contreras Pe{\~n}a}, \& {Johnstone}}]{2021Lee}
{Lee}, J.-E., {Lee}, S., {Lee}, S., {et~al.} 2021, \bibinfo{title}{{AGB Interlopers in YSO Catalogs Hunted out by NEOWISE},} \apjl, 916, L20, \dodoi{10.3847/2041-8213/ac0d59}

\bibitem[{S. {Lee} {et~al.}(2024){Lee}, {Lee}, {Contreras Pe{\~n}a}, {Johnstone}, {Herczeg}, \& {Lee}}]{2024LeeS}
{Lee}, S., {Lee}, J.-E., {Contreras Pe{\~n}a}, C., {et~al.} 2024, \bibinfo{title}{{Mid-infrared Variability of Young Stellar Objects on Timescales of Days to Years},} \apj, 962, 38, \dodoi{10.3847/1538-4357/ad14f8}

\bibitem[{Y.-H. {Lee} {et~al.}(2021){Lee}, {Johnstone}, {Lee}, {Herczeg}, {Mairs}, {Contreras-Pe{\~n}a}, {Hatchell}, {Naylor}, {Bell}, {Bourke}, {Broughton}, {Francis}, {Gupta}, {Harsono}, {Liu}, {Park}, {Plovie}, {Moriarty-Schieven}, {Scholz}, {Sharma}, {Teixeira}, {Wang}, {Aikawa}, {Bower}, {Vivien Chen}, {Bae}, {Baek}, {Chapman}, {Ping Chen}, {Du}, {Dutta}, {Forbrich}, {Guo}, {Inutsuka}, {Kang}, {Kirk}, {Kuan}, {Kwon}, {Lai}, {Lalchand}, {Lane}, {Lee}, {Liu}, {Morata}, {Pearson}, {Pon}, {Sahu}, {Shang}, {Stamatellos}, {Tang}, {Xu}, {Yoo}, \& {Rawlings}}]{2021LeeY}
{Lee}, Y.-H., {Johnstone}, D., {Lee}, J.-E., {et~al.} 2021, \bibinfo{title}{{The JCMT Transient Survey: Four-year Summary of Monitoring the Submillimeter Variability of Protostars},} \apj, 920, 119, \dodoi{10.3847/1538-4357/ac1679}

\bibitem[{N.~R. {Lomb}(1976){Lomb}}]{1976Lomb}
{Lomb}, N.~R. 1976, \bibinfo{title}{{Least-Squares Frequency Analysis of Unequally Spaced Data},} \apss, 39, 447, \dodoi{10.1007/BF00648343}

\bibitem[{S.~L. {Lumsden} {et~al.}(2002){Lumsden}, {Hoare}, {Oudmaijer}, \& {Richards}}]{2002Lumsden}
{Lumsden}, S.~L., {Hoare}, M.~G., {Oudmaijer}, R.~D., \& {Richards}, D. 2002, \bibinfo{title}{{The population of the Galactic plane as seen by MSX},} \mnras, 336, 621, \dodoi{10.1046/j.1365-8711.2002.05785.x}

\bibitem[{A. {Mainzer} {et~al.}(2014){Mainzer}, {Bauer}, {Cutri}, {Grav}, {Masiero}, {Beck}, {Clarkson}, {Conrow}, {Dailey}, {Eisenhardt}, {Fabinsky}, {Fajardo-Acosta}, {Fowler}, {Gelino}, {Grillmair}, {Heinrichsen}, {Kendall}, {Kirkpatrick}, {Liu}, {Masci}, {McCallon}, {Nugent}, {Papin}, {Rice}, {Royer}, {Ryan}, {Sevilla}, {Sonnett}, {Stevenson}, {Thompson}, {Wheelock}, {Wiemer}, {Wittman}, {Wright}, \& {Yan}}]{2014Mainer}
{Mainzer}, A., {Bauer}, J., {Cutri}, R.~M., {et~al.} 2014, \bibinfo{title}{{Initial Performance of the NEOWISE Reactivation Mission},} \apj, 792, 30, \dodoi{10.1088/0004-637X/792/1/30}

\bibitem[{G. {Marton} {et~al.}(2016){Marton}, {T{\'o}th}, {Paladini}, {Kun}, {Zahorecz}, {McGehee}, \& {Kiss}}]{2016Marton}
{Marton}, G., {T{\'o}th}, L.~V., {Paladini}, R., {et~al.} 2016, \bibinfo{title}{{An all-sky support vector machine selection of WISE YSO candidates},} \mnras, 458, 3479, \dodoi{10.1093/mnras/stw398}

\bibitem[{G. {Marton} {et~al.}(2019){Marton}, {{\'A}brah{\'a}m}, {Szegedi-Elek}, {Varga}, {Kun}, {K{\'o}sp{\'a}l}, {Varga-Vereb{\'e}lyi}, {Hodgkin}, {Szabados}, {Beck}, \& {Kiss}}]{2019Marton}
{Marton}, G., {{\'A}brah{\'a}m}, P., {Szegedi-Elek}, E., {et~al.} 2019, \bibinfo{title}{{Identification of Young Stellar Object candidates in the Gaia DR2 x AllWISE catalogue with machine learning methods},} \mnras, 487, 2522, \dodoi{10.1093/mnras/stz1301}

\bibitem[{M. {Matsuura} {et~al.}(2006){Matsuura}, {Wood}, {Sloan}, {Zijlstra}, {van Loon}, {Groenewegen}, {Blommaert}, {Cioni}, {Feast}, {Habing}, {Hony}, {Lagadec}, {Loup}, {Menzies}, {Waters}, \& {Whitelock}}]{2006Matsuura}
{Matsuura}, M., {Wood}, P.~R., {Sloan}, G.~C., {et~al.} 2006, \bibinfo{title}{{Spitzer observations of acetylene bands in carbon-rich asymptotic giant branch stars in the Large Magellanic Cloud},} \mnras, 371, 415, \dodoi{10.1111/j.1365-2966.2006.10664.x}

\bibitem[{S.~T. {Megeath} {et~al.}(2012){Megeath}, {Gutermuth}, {Muzerolle}, {Kryukova}, {Flaherty}, {Hora}, {Allen}, {Hartmann}, {Myers}, {Pipher}, {Stauffer}, {Young}, \& {Fazio}}]{2012Megeath}
{Megeath}, S.~T., {Gutermuth}, R., {Muzerolle}, J., {et~al.} 2012, \bibinfo{title}{{The Spitzer Space Telescope Survey of the Orion A and B Molecular Clouds. I. A Census of Dusty Young Stellar Objects and a Study of Their Mid-infrared Variability},} \aj, 144, 192, \dodoi{10.1088/0004-6256/144/6/192}

\bibitem[{M.~M. {Miller Bertolami}(2016){Miller Bertolami}}]{2016Miguel}
{Miller Bertolami}, M.~M. 2016, \bibinfo{title}{{New models for the evolution of post-asymptotic giant branch stars and central stars of planetary nebulae},} \aap, 588, A25, \dodoi{10.1051/0004-6361/201526577}

\bibitem[{F. Murtagh(1991)Murtagh}]{1991Murtagh}
Murtagh, F. 1991, \bibinfo{title}{Multilayer perceptrons for classification and regression,} Neurocomputing, 2, 183, \dodoi{https://doi.org/10.1016/0925-2312(91)90023-5}

\bibitem[{L.~A. {Nyman} {et~al.}(1998){Nyman}, {Hall}, \& {Olofsson}}]{1998Nyman}
{Nyman}, L.~A., {Hall}, P.~J., \& {Olofsson}, H. 1998, \bibinfo{title}{{SiO masers in OH/IR stars, proto-planetary and planetary nebulae},} \aaps, 127, 185, \dodoi{10.1051/aas:1998343}

\bibitem[{K. {O'Shea} \& R. {Nash}(2015){O'Shea} \& {Nash}}]{2015CNN}
{O'Shea}, K., \& {Nash}, R. 2015, \bibinfo{title}{{An Introduction to Convolutional Neural Networks},} arXiv e-prints, arXiv:1511.08458, \dodoi{10.48550/arXiv.1511.08458}

\bibitem[{W. {Park} {et~al.}(2021){Park}, {Lee}, {Contreras Pe{\~n}a}, {Johnstone}, {Herczeg}, {Lee}, {Lee}, {Bhardwaj}, \& {Moriarty-Schieven}}]{2021Park}
{Park}, W., {Lee}, J.-E., {Contreras Pe{\~n}a}, C., {et~al.} 2021, \bibinfo{title}{{Quantifying Variability of Young Stellar Objects in the Mid-infrared Over 6 Years with the Near-Earth Object Wide-field Infrared Survey Explorer},} \apj, 920, 132, \dodoi{10.3847/1538-4357/ac1745}

\bibitem[{A. {Paszke} {et~al.}(2019){Paszke}, {Gross}, {Massa}, {Lerer}, {Bradbury}, {Chanan}, {Killeen}, {Lin}, {Gimelshein}, {Antiga}, {Desmaison}, {K{\"o}pf}, {Yang}, {DeVito}, {Raison}, {Tejani}, {Chilamkurthy}, {Steiner}, {Fang}, {Bai}, \& {Chintala}}]{2019Pytorch}
{Paszke}, A., {Gross}, S., {Massa}, F., {et~al.} 2019, \bibinfo{title}{{PyTorch: An Imperative Style, High-Performance Deep Learning Library},} arXiv e-prints, arXiv:1912.01703, \dodoi{10.48550/arXiv.1912.01703}

\bibitem[{F. Pedregosa {et~al.}(2011)Pedregosa, Varoquaux, Gramfort, Michel, Thirion, Grisel, Blondel, Prettenhofer, Weiss, Dubourg, Vanderplas, Passos, Cournapeau, Brucher, Perrot, \& Duchesnay}]{2011scikit-learn}
Pedregosa, F., Varoquaux, G., Gramfort, A., {et~al.} 2011, \bibinfo{title}{Scikit-learn: Machine Learning in {P}ython,} Journal of Machine Learning Research, 12, 2825

\bibitem[{L. {Perez} \& J. {Wang}(2017){Perez} \& {Wang}}]{2017DataAugmentation}
{Perez}, L., \& {Wang}, J. 2017, \bibinfo{title}{{The Effectiveness of Data Augmentation in Image Classification using Deep Learning},} arXiv e-prints, arXiv:1712.04621, \dodoi{10.48550/arXiv.1712.04621}

\bibitem[{T.~P. {Robitaille} {et~al.}(2008){Robitaille}, {Meade}, {Babler}, {Whitney}, {Johnston}, {Indebetouw}, {Cohen}, {Povich}, {Sewilo}, {Benjamin}, \& {Churchwell}}]{2008Robit}
{Robitaille}, T.~P., {Meade}, M.~R., {Babler}, B.~L., {et~al.} 2008, \bibinfo{title}{{Intrinsically Red Sources Observed by Spitzer in the Galactic Midplane},} \aj, 136, 2413, \dodoi{10.1088/0004-6256/136/6/2413}

\bibitem[{G.~A. {Romero} {et~al.}(2012){Romero}, {Schreiber}, {Cieza}, {Rebassa-Mansergas}, {Mer{\'\i}n}, {Smith Castelli}, {Allen}, \& {Morrell}}]{2012Romero}
{Romero}, G.~A., {Schreiber}, M.~R., {Cieza}, L.~A., {et~al.} 2012, \bibinfo{title}{{The Nature of Transition Circumstellar Disks. II. Southern Molecular Clouds},} \apj, 749, 79, \dodoi{10.1088/0004-637X/749/1/79}

\bibitem[{J.~D. {Scargle}(1989){Scargle}}]{1989Scargle}
{Scargle}, J.~D. 1989, \bibinfo{title}{{Studies in Astronomical Time Series Analysis. III. Fourier Transforms, Autocorrelation Functions, and Cross-Correlation Functions of Unevenly Spaced Data},} \apj, 343, 874, \dodoi{10.1086/167757}

\bibitem[{R.~R. {Selvaraju} {et~al.}(2016){Selvaraju}, {Cogswell}, {Das}, {Vedantam}, {Parikh}, \& {Batra}}]{2016Gradcam}
{Selvaraju}, R.~R., {Cogswell}, M., {Das}, A., {et~al.} 2016, \bibinfo{title}{{Grad-CAM: Visual Explanations from Deep Networks via Gradient-based Localization},} arXiv e-prints, arXiv:1610.02391, \dodoi{10.48550/arXiv.1610.02391}

\bibitem[{A. {Sherstinsky}(2020){Sherstinsky}}]{2020RNN_funda}
{Sherstinsky}, A. 2020, \bibinfo{title}{{Fundamentals of Recurrent Neural Network (RNN) and Long Short-Term Memory (LSTM) network},} Physica D Nonlinear Phenomena, 404, 132306, \dodoi{10.1016/j.physd.2019.132306}

\bibitem[{M.~F. {Skrutskie} {et~al.}(2006){Skrutskie}, {Cutri}, {Stiening}, {Weinberg}, {Schneider}, {Carpenter}, {Beichman}, {Capps}, {Chester}, {Elias}, {Huchra}, {Liebert}, {Lonsdale}, {Monet}, {Price}, {Seitzer}, {Jarrett}, {Kirkpatrick}, {Gizis}, {Howard}, {Evans}, {Fowler}, {Fullmer}, {Hurt}, {Light}, {Kopan}, {Marsh}, {McCallon}, {Tam}, {Van Dyk}, \& {Wheelock}}]{20062MASS}
{Skrutskie}, M.~F., {Cutri}, R.~M., {Stiening}, R., {et~al.} 2006, \bibinfo{title}{{The Two Micron All Sky Survey (2MASS)},} \aj, 131, 1163, \dodoi{10.1086/498708}

\bibitem[{N. Srivastava {et~al.}(2014)Srivastava, Hinton, Krizhevsky, Sutskever, \& Salakhutdinov}]{2014Dropout}
Srivastava, N., Hinton, G., Krizhevsky, A., Sutskever, I., \& Salakhutdinov, R. 2014, \bibinfo{title}{Dropout: A Simple Way to Prevent Neural Networks from Overfitting,} Journal of Machine Learning Research, 15, 1929.
\newblock \url{http://jmlr.org/papers/v15/srivastava14a.html}

\bibitem[{K.-W. {Suh}(2021){Suh}}]{2021Suh}
{Suh}, K.-W. 2021, \bibinfo{title}{{A New Catalog of Asymptotic Giant Branch Stars in Our Galaxy},} \apjs, 256, 43, \dodoi{10.3847/1538-4365/ac1274}

\bibitem[{K.-W. {Suh}(2024){Suh}}]{2024Suh}
{Suh}, K.-W. 2024, \bibinfo{title}{{AGB and Post-AGB Stars versus Planetary Nebulae and Young Stellar Objects: Properties in Visual and IR Bands},} arXiv e-prints, arXiv:2407.20540, \dodoi{10.48550/arXiv.2407.20540}

\bibitem[{A.~M. {Tatarnikov} {et~al.}(2024){Tatarnikov}, {Zheltoukhov}, \& {Malik}}]{2024Tatarnikov}
{Tatarnikov}, A.~M., {Zheltoukhov}, S.~G., \& {Malik}, E.~D. 2024, \bibinfo{title}{{Spectral Energy Distribution of Late Stage Stars},} Moscow University Physics Bulletin, 79, 385, \dodoi{10.3103/S0027134924700425}

\bibitem[{M.~M. Taye(2023)Taye}]{2023OverallMLDL}
Taye, M.~M. 2023, \bibinfo{title}{Understanding of Machine Learning with Deep Learning: Architectures, Workflow, Applications and Future Directions,} Computers, 12, \dodoi{10.3390/computers12050091}

\bibitem[{W.~E.~C.~J. {van der Veen} \& H.~J. {Habing}(1988){van der Veen} \& {Habing}}]{1988Vanderveen}
{van der Veen}, W.~E.~C.~J., \& {Habing}, H.~J. 1988, \bibinfo{title}{{The IRAS two-colour diagram as a tool for studying late stages of stellar evolution.},} \aap, 194, 125

\bibitem[{J.~T. {van Loon} {et~al.}(1999){van Loon}, {Groenewegen}, {de Koter}, {Trams}, {Waters}, {Zijlstra}, {Whitelock}, \& {Loup}}]{1999VanLoon}
{van Loon}, J.~T., {Groenewegen}, M.~A.~T., {de Koter}, A., {et~al.} 1999, \bibinfo{title}{{Mass-loss rates and luminosity functions of dust-enshrouded AGB stars and red supergiants in the LMC},} \aap, 351, 559, \dodoi{10.48550/arXiv.astro-ph/9909416}

\bibitem[{A. {Vaswani} {et~al.}(2017){Vaswani}, {Shazeer}, {Parmar}, {Uszkoreit}, {Jones}, {Gomez}, {Kaiser}, \& {Polosukhin}}]{2017Transformer}
{Vaswani}, A., {Shazeer}, N., {Parmar}, N., {et~al.} 2017, \bibinfo{title}{{Attention Is All You Need},} arXiv e-prints, arXiv:1706.03762, \dodoi{10.48550/arXiv.1706.03762}

\bibitem[{V. {Venkata Raman} \& B.~G. {Anandarao}(2008){Venkata Raman} \& {Anandarao}}]{2008Raman}
{Venkata Raman}, V., \& {Anandarao}, B.~G. 2008, \bibinfo{title}{{Infrared spectroscopic study of a selection of AGB and post-AGB stars},} \mnras, 385, 1076, \dodoi{10.1111/j.1365-2966.2008.12915.x}

\bibitem[{E.~I. {Vorobyov} \& S. {Basu}(2015){Vorobyov} \& {Basu}}]{2015Vorobyov}
{Vorobyov}, E.~I., \& {Basu}, S. 2015, \bibinfo{title}{{Variable Protostellar Accretion with Episodic Bursts},} \apj, 805, 115, \dodoi{10.1088/0004-637X/805/2/115}

\bibitem[{L.~J. {Whitehouse} {et~al.}(2021){Whitehouse}, {Farihi}, {Howarth}, {Mancino}, {Walters}, {Swan}, {Wilson}, \& {Guo}}]{2021Whitehouse}
{Whitehouse}, L.~J., {Farihi}, J., {Howarth}, I.~D., {et~al.} 2021, \bibinfo{title}{{Carbon-enhanced stars with short orbital and spin periods},} \mnras, 506, 4877, \dodoi{10.1093/mnras/stab1913}

\bibitem[{P.~A. {Whitelock} {et~al.}(2008){Whitelock}, {Feast}, \& {Van Leeuwen}}]{2008Whitelock}
{Whitelock}, P.~A., {Feast}, M.~W., \& {Van Leeuwen}, F. 2008, \bibinfo{title}{{AGB variables and the Mira period-luminosity relation},} \mnras, 386, 313, \dodoi{10.1111/j.1365-2966.2008.13032.x}

\bibitem[{J.~P. {Williams} \& L.~A. {Cieza}(2011){Williams} \& {Cieza}}]{2011Williams}
{Williams}, J.~P., \& {Cieza}, L.~A. 2011, \bibinfo{title}{{Protoplanetary Disks and Their Evolution},} \araa, 49, 67, \dodoi{10.1146/annurev-astro-081710-102548}

\bibitem[{E.~L. {Wright} {et~al.}(2010){Wright}, {Eisenhardt}, {Mainzer}, {Ressler}, {Cutri}, {Jarrett}, {Kirkpatrick}, {Padgett}, {McMillan}, {Skrutskie}, {Stanford}, {Cohen}, {Walker}, {Mather}, {Leisawitz}, {Gautier}, {McLean}, {Benford}, {Lonsdale}, {Blain}, {Mendez}, {Irace}, {Duval}, {Liu}, {Royer}, {Heinrichsen}, {Howard}, {Shannon}, {Kendall}, {Walsh}, {Larsen}, {Cardon}, {Schick}, {Schwalm}, {Abid}, {Fabinsky}, {Naes}, \& {Tsai}}]{2010WISE}
{Wright}, E.~L., {Eisenhardt}, P. R.~M., {Mainzer}, A.~K., {et~al.} 2010, \bibinfo{title}{{The Wide-field Infrared Survey Explorer (WISE): Mission Description and Initial On-orbit Performance},} \aj, 140, 1868, \dodoi{10.1088/0004-6256/140/6/1868}

\bibitem[{N.~J. {Wright} {et~al.}(2009){Wright}, {Barlow}, {Greimel}, {Drew}, {Matsuura}, {Unruh}, \& {Zijlstra}}]{2009Wright}
{Wright}, N.~J., {Barlow}, M.~J., {Greimel}, R., {et~al.} 2009, \bibinfo{title}{{Near-IR spectra of IPHAS extremely red Galactic AGB stars},} \mnras, 400, 1413, \dodoi{10.1111/j.1365-2966.2009.15536.x}

\bibitem[{G. {Ye} {et~al.}(2024){Ye}, {Zhang}, \& {Wu}}]{2024Ye}
{Ye}, G., {Zhang}, H., \& {Wu}, Q. 2024, \bibinfo{title}{{Machine Learning-based Search of High-redshift Quasars},} arXiv e-prints, arXiv:2409.02167, \dodoi{10.48550/arXiv.2409.02167}

\bibitem[{K. Zantvoort {et~al.}(2024)Zantvoort, Nacke, G{\"o}rlich, Hornstein, Jacobi, \& Funk}]{2024Numtestset}
Zantvoort, K., Nacke, B., G{\"o}rlich, D., {et~al.} 2024, \bibinfo{title}{Estimation of minimal data sets sizes for machine learning predictions in digital mental health interventions,} NPJ Digit. Med., 7, 361

\end{thebibliography}

%\begin{comment}

\appendix
\restartappendixnumbering
\section{Technical Interpretation}

\begin{figure*}
    \centering
    \includegraphics[width=0.9\textwidth]{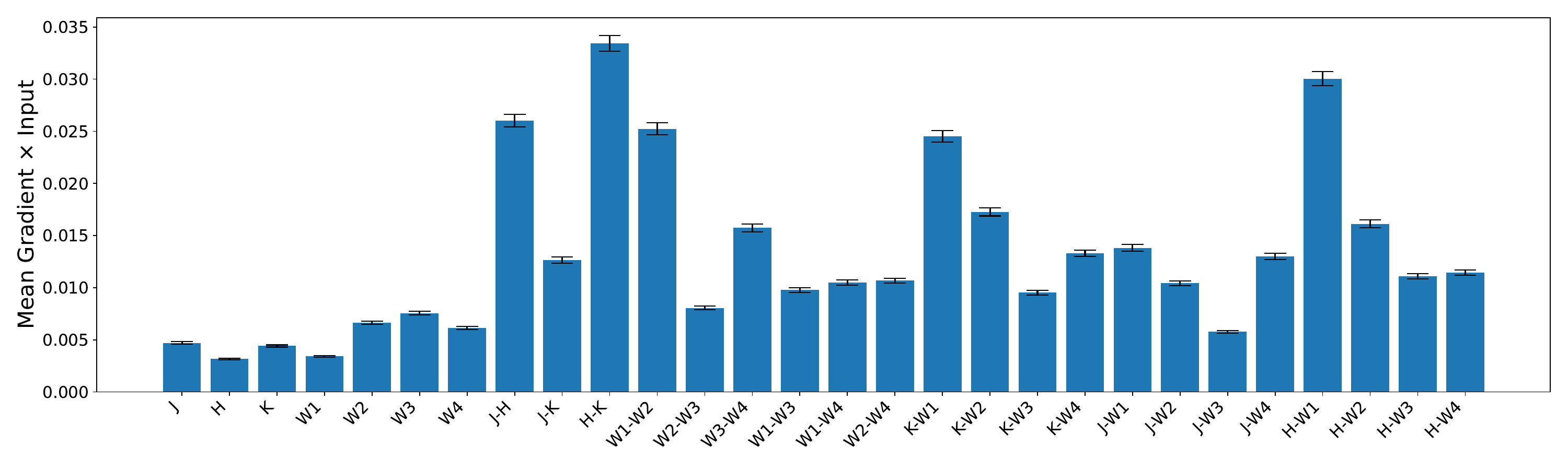}
    \includegraphics[width=0.3\textwidth ]{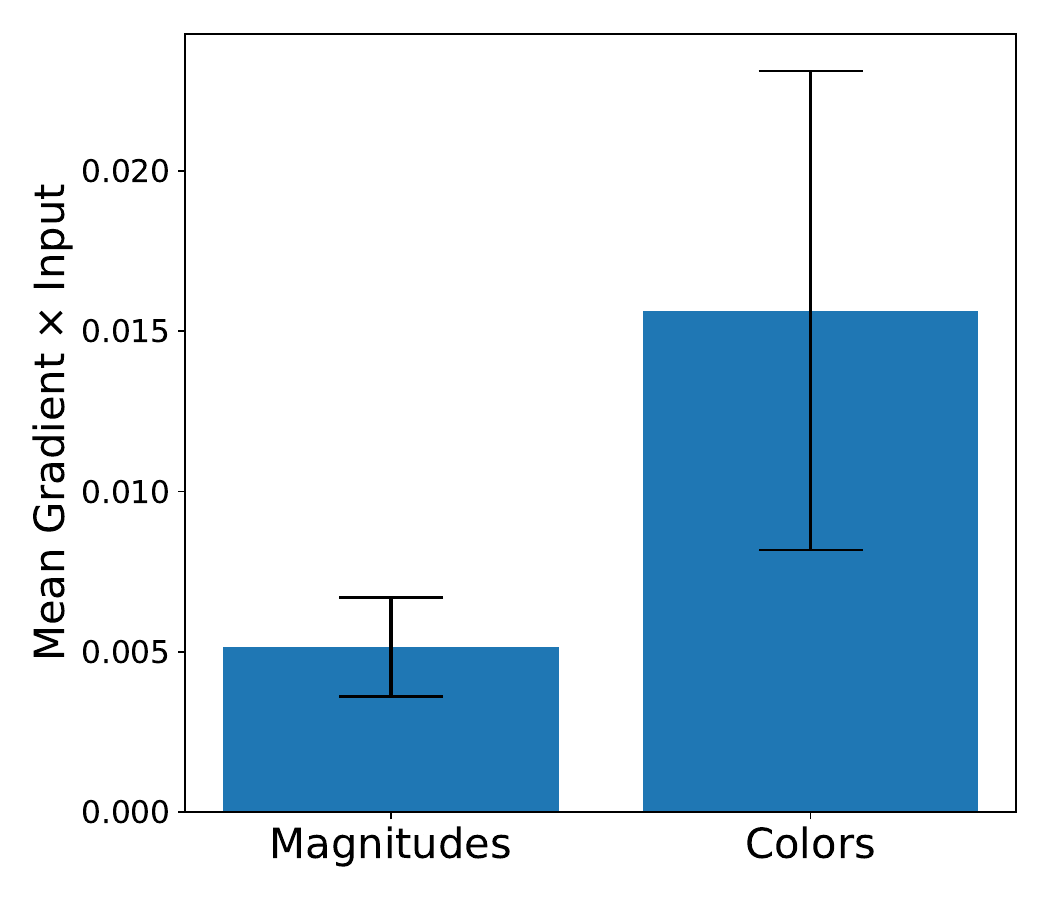}
    \includegraphics[width=0.6\textwidth ]{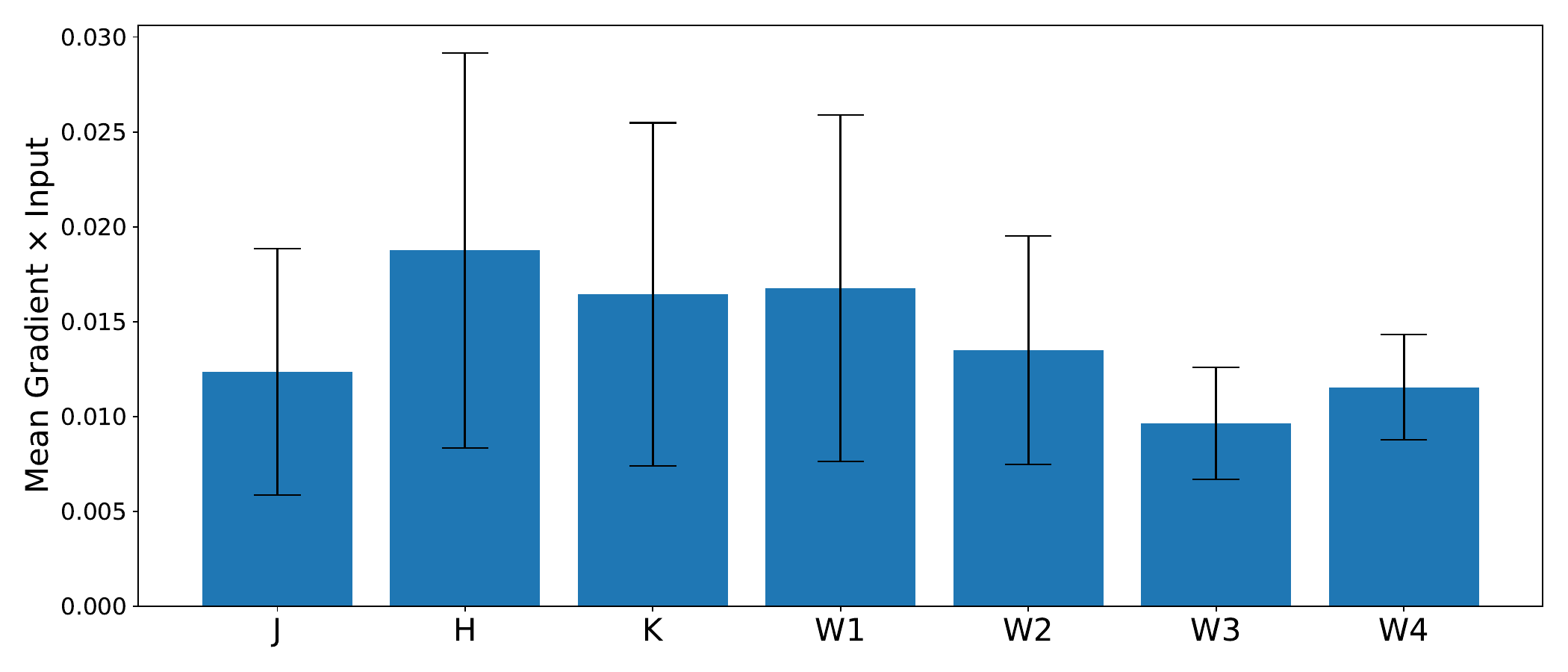}

    \flushleft{\textbf{Notes}. The error bars represent the standard deviation of each histogram.}
    
    \caption{\textit{Top}: The histogram of the 28 input features of MLP in Filter 1. \textit{Bottom left}: The histogram of the input features grouped by ``Magnitudes" and ``Colors". \textit{Bottom right}: The histogram grouped by associated 2MASS and AllWISE filters.}
    
    \label{figAp1}
\end{figure*}

\begin{figure}
    \centering
    \includegraphics[width=0.45\textwidth ]{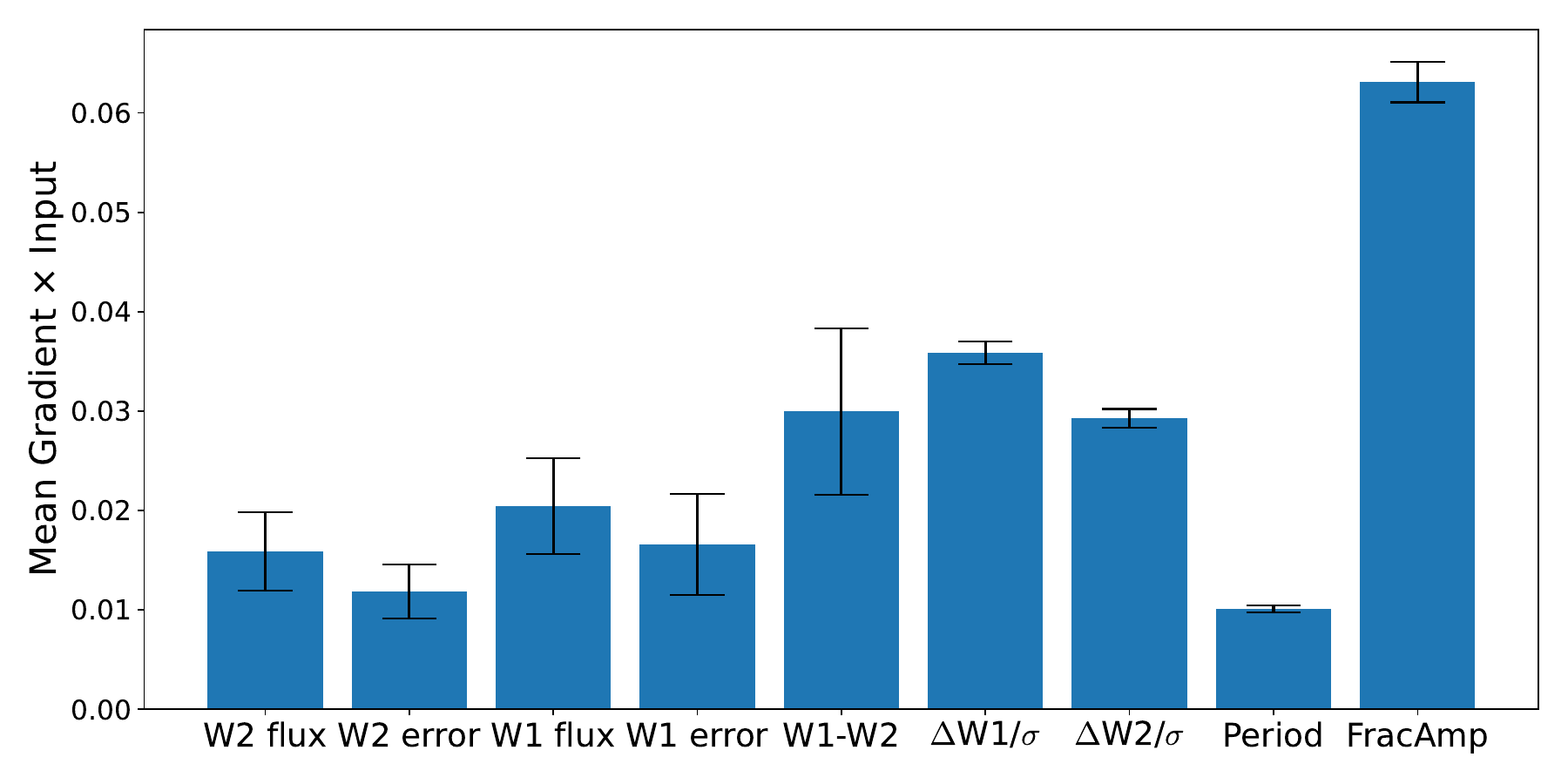}
    \includegraphics[width=0.45\textwidth ]{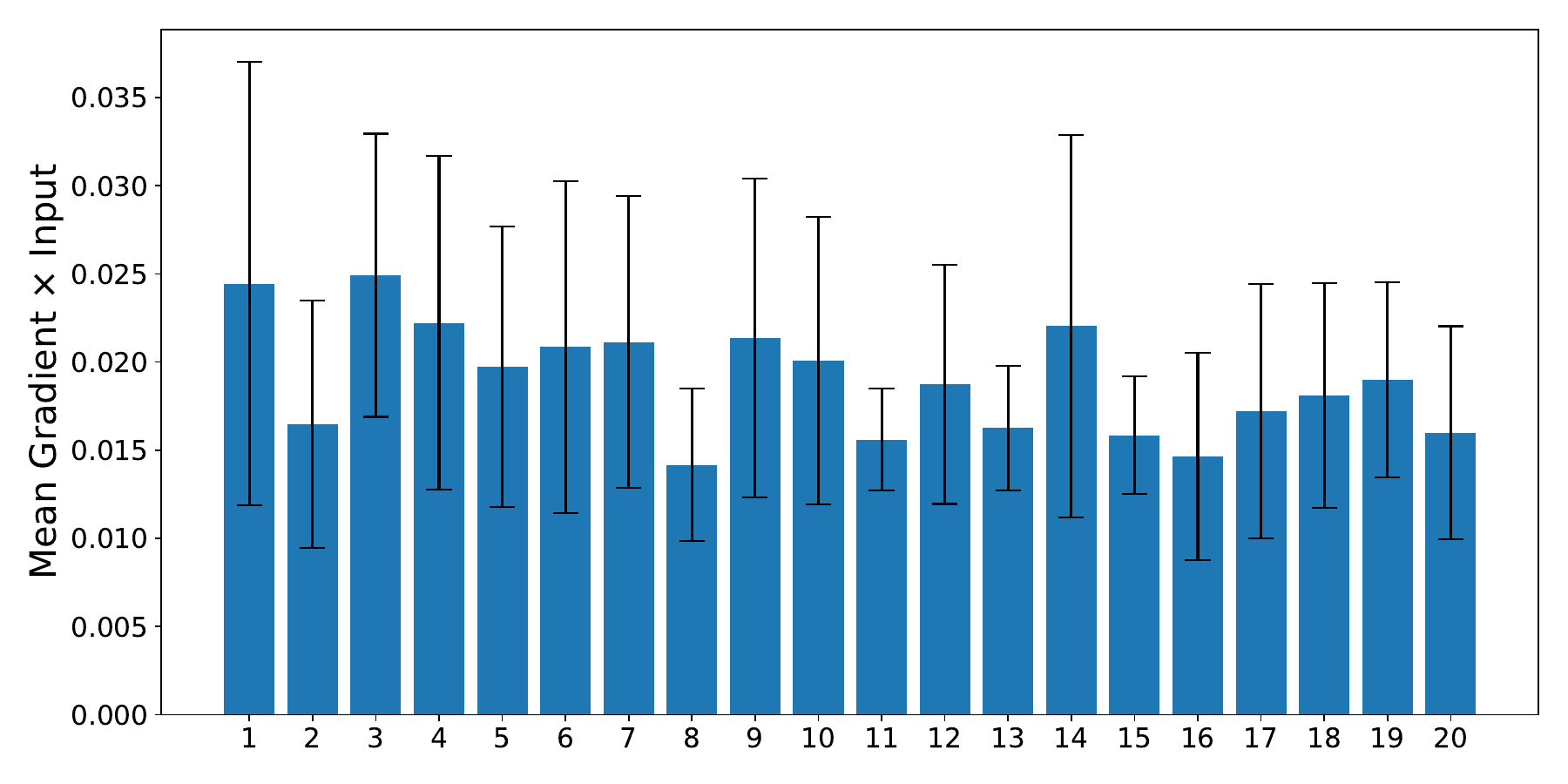}
    
    \caption{The histograms of the input features on MLP in Filter 2. \textit{Top}: The histogram categorized by characteristics. \textit{Bottom}: The histogram categorized by observational epochs by numbers.}
    \label{figAp2}
\end{figure}

{
%Appendix A 수정사항: Delta W1 /sigma 는 fractional amplitude의 의미와 같으며, 값이 클 수록 AGB stars이고 작을수록 YSOs라고 본다. fractional Amplitude와의 차이점은, frac-Amp은 false alarm prob. 가 0.01 보다 작을 경우에만 값을 갖기 때문에, 주로 periodic 한 lightcurve를 보이는 YSO or AGB에서 분류하는데 도움이 되므로 이런 변광특징을 보이는 AGBs는 frac-Amp and Delta W1/W2 값 모두를 활용하고, 이 외 변광 특징이 크지 않은 애들의 경우, Delta W1 /W2가 AGB or YSO 분류에 도움을 주기 때문에 (이유는 맨 위의 내용) 모델이 분류하는데 이 세 값들을 모두 고르게 보는 것으로 추정된다. 

Despite the difficulties in interpreting the ML process, the Grad-CAM method \citep{2016Gradcam} provides one restricted ML analysis. This method uses gradient values in MLP to trace back and compare ``relatively" which input data is taken into account the most. We analyzed our MLPs only in Filters 1 and 2 because SVM and RF do not apply to Grad-CAM. Instead, SVM and RF are interpretable compared to the MLP-based models, and they followed the general classification trends based on CMD and CCD from existing studies \citep{2008Robit, 2014K&L}.

Since Grad-CAM is predominantly designed for 2D or 3D image data, we modified Grad-CAM to be applicable to 1D data to fit our model, multiplying the gradients from backpropagation and the training input data values. Figure \ref{figAp1} shows the Grad-CAM result of the MLP in Filter 1 with various input groupings as histograms. The model seems to be sensitive to color indices rather than magnitudes, especially H-K, J-H, K-W1, H-W1, and W1-W2. These trends are apparent when we grouped the inputs into ``Magnitudes" and ``Colors", where Magnitudes cover 2MASS J, H, K and AllWISE W1, W2, W3, W4 values, and Colors consider all the color combinations with 2MASS and AllWISE values (bottom left histogram of Figure \ref{figAp1}). 

The bottom right histogram of Figure \ref{figAp1} represents filter-related groups (e.g., the J magnitude data goes to the `J' group, while the H-K color index belongs to both `H' and `K' groups). No significant differences were found between the filter groups. Therefore, we estimate that the MLP in the Filter 1 model prefers to judge color indices rather than magnitudes, considering each IR filter evenly. This aligns with the existing analysis, which indicates that color is a more reliable indicator than magnitude, and the H-K color index is the most effective in classifying YSOs and AGB stars \citep{2008Robit}. 

Figure \ref{figAp2} shows the Grad-CAM results of the MLP classifier in Filter 2. We grouped the 104 input features in Filter 2 by their characteristics, namely W1 fluxes, W1 errors, W2 fluxes, W2 errors, $\Delta$W1/$\sigma$, $\Delta$W2/$\sigma$, Period, and fAMP. The top histogram of Figure \ref{figAp2} illustrates that Filter 2 is sensitive to amplitude-related information: fAMP, $\Delta$W1/$\sigma$, and $\Delta$W2/$\sigma$. fAMP for {\bf``periodic"} YSOs and AGB stars, and $\Delta$W1/$\sigma$ and $\Delta$W2/$\sigma$ for all sources regardless of their periodicity. It is explicable that light curves in AGB stars show a large amplitude due to {\bf stellar pulsation} \citep{2008Whitelock}. YSOs, on the other hand, reveal a small amplitude because periodic YSOs are uncommon and their periodic mechanism are mostly due to disk rotation \citep{2021Guo}. In contrast, Period is no longer a key characteristic once periodic variability has been confirmed. Finally, it is predictable that the model considers W1-W2 input because the colors of AGB stars are generally bluer than those of YSOs \citep{2008Robit, 2014K&L}. 

The bottom histogram of Figure \ref{figAp2} shows the dependency of the Filter 2 model for the individual NEOWISE epochs. The model does not appear to be biased toward any specific epoch. This outcome is expected, as AGB stars and YSOs should not show any preference for particular observation times. Thus, these results suggest that the Filter 2 model was properly trained. 

Based on the technical analysis using Grad-CAM, we suggest that MLPs in Filters 1 and 2 are well-trained. The model accounts for differences between AGB stars and YSOs in magnitude, color, and light-curve trends. However, since the Grad-CAM method merely indicates which inputs are important, not why, the above analysis is an interpretation based on the correlation between Grad-CAM results and scientific approaches.
}

\section{Ensemble Structure of Filter 1} \label{appB}

\begin{table}[!hbt]
    \caption{The pseudo-code for the customized ensemble method of Filter 1.}
    \label{tabap1}
    \centering
    \subfloat{\fbox{\resizebox{\columnwidth}{!}{\begin{tabular}{l} %
           Class Ensemble\_model():\\
           \\
               def\_\_init\_\_(self, estimators, voting):\\ 
                   self.estimators = estimators\\
                   self.voting = voting\\
            \\    
                def predict(self, X\_input):\\
                    X\_pred = [estimators.pred(X\_input)]\\
            \\
                    \texttt{if self.voting == `soft'}:\\
                        YSO probability = sum(X\_pred[:, ::2], axis=1) / (number of estimators)\\
                        AGB probability = sum(X\_pred[:, 1::2], axis=1) / (number of estimators)\\
            \\
                        \texttt{for i in range(X\_pred.shape[0])}:\\
                            soft\_vote = append([YSO probability, AGB probability]) (N $\times$ 2 array)\\
                        return soft\_vote\\
            \\
                    \texttt{if self.voting == `hard'}:\\
                        \texttt{for i in range(X\_pred.shape[1]//2)}:\\
                            np.argmax(np.column\_stack([X\_pred[:, 2 $\times$ i], X\_pred[:, 2 $\times$ i+1]]), axis=1)\\
                        most\_frequent\_values = np.apply\_along\_axis(lambda x: np.bincount(x).argmax(), axis=1)\\
            \\
                        return most\_frequent\_values\\ 

    \end{tabular}% 
    }}}
\end{table}

The voting ensemble method is the main structure of Filter 1 from our model. The \texttt{VotingClassifier} ensemble method in the \texttt{scikit-learn} package is widely used for voting ensemble. Since it is only compatible with ML methods within \texttt{scikit-learn}, we customized the ensemble method to handle ML methods from both \texttt{scikit-learn} and \texttt{pytorch}. The pseudo code for our ensemble structure is shown in Table \ref{tabap1}. As we use SVM, RF, and MLP for the photometry classification, we set \texttt{estimators = [`SVM,' `RF,' `MLP']}.

Hard and soft voting are often referred to as majority voting and averaging voting \citep{2021Ma_Ensemble_Review}. Hard voting is selecting the most frequent classifier result, as represented in the equation below, where $\hat{y_i}$ is the prediction of the $i$-th ML method.

\begin{equation}
Mode(\hat{y_1}, \hat{y_2}, \hat{y_3}, ... ,\hat{y_i})
\end{equation}
Soft voting, on the other hand, averages the final classification probabilities ($\hat{p_i}$) among the ML methods. 
\begin{equation}
argmax(\frac{1}{n}\sum^{n}{\hat{p_i}})
\end{equation}

\section{AGB candidates catalog} \label{appC}
\restartappendixnumbering

 We provide a catalog for the 258 AGB candidates from the SPICY catalog obtained with the Double Filter model. Table \ref{tab_258_cand} provides the column information of the catalog, and the catalog is given in Table \ref{tab_258_cand_mr}.

\begin{table*}[!hbt]
\centering
\caption{Column information for the 258 AGB candidates.}
\label{tab_258_cand}
\begin{tabularx}{\linewidth}{@{\extracolsep{\fill}}llX @{}}
\toprule
{\bf Column} & {\bf Column Name} & {\bf Description (unit)} \\ 
\hline
1 & ID  & SPICY ID from \citet{2021Kuhn} \\
2 & RAJ2000  & RA (deg)\\
3 & DEJ2000 & DEC (deg)\\
4 - 23 & w2t1 $\sim$ w2t20 & NEOWISE W2 flux (Jy)\\ 
24 - 43 & w2te1 $\sim$ w2te20 & NEOWISE W2 flux error (Jy)\\
44 - 63 & w1t1 $\sim$ w1t20 & NEOWISE W1 flux (Jy)\\
64 - 83 & w1te1 $\sim$ w1te20 & NEOWISE W1 flux error (Jy)\\
84 - 103 & w1t1-w2t1 $\sim$ w1t20-w2t20 & NEOWISE W1-W2 color (mag)\\
104 & DelW1 & $\Delta$W1/$\sigma$ from Eq. \ref{eq1-1}\\
105 & DelW2 & $\Delta$W2/$\sigma$ from Eq. \ref{eq1-2}\\
106 & Period & Normalized period calculated by Lomb-Scargle periodogram. Zeros for non-periodic sources\\
107 & fracAmp & Normalized fractional amplitude calculated with Eq. \ref{eq2}. Zeros for non-periodic sources\\ \hline
108 & J   & 2MASS J magnitude (mag)\\
109 & H   & 2MASS H magnitude (mag)\\
110 & K   & 2MASS K magnitude (mag)\\
111 & W1  & AllWISE W1 magnitude (mag)\\
112 & W2  & AllWISE W2 magnitude (mag)\\
113 & W3  & AllWISE W3 magnitude (mag)\\
114 & W4  & AllWISE W4 magnitude (mag)\\
115 - 135 & J-H, H-K, ..., H-W4  & Color combinations of 2MASS and AllWISE (mag)\\
\hline

\end{tabularx}
\end{table*}

\begin{table*}[!hbt]
\centering
\caption{Catalog of 258 AGB candidates from SPICY obtained with the Double Filter model.}
\label{tab_258_cand_mr}
\begin{threeparttable}
\begin{tabularx}{\linewidth}{@{\extracolsep{\fill}} XXXXXXXXX @{}}
\toprule
{\bf ID} & {\bf RAJ2000} & {\bf DEJ2000} & {\bf w2t1} & {\bf w2t2} & {\bf ...} & {\bf H-W2} & {\bf H-W3} & {\bf H-W4} \\ 
\hline

4915 & 150.0483 & -56.2529 & 0.03258 & 0.03264 & ... & 1.975 & 2.603 & 2.806\\
... & ... & ... & ... & ... & ... & ... & ... & ...\\
116666 & 337.3901 & 60.9613 & 0.00018 & 0.00018 & ... & -0.070 & 2.037 & 5.748\\
\hline
\end{tabularx} \vspace{5pt}
    \begin{tablenotes}[flushleft]\footnotesize
        \item[] (This table is available in its entirety in machine-readable form.)
        \end{tablenotes}
\end{threeparttable}
\end{table*}

\end{document}